\documentclass[10pt,preprint]{aastex}

\newcommand{\lya}{Ly{\sc $\alpha$}}
\newcommand{\lyal}{Ly{\sc $\alpha$}\,$\lambda$1216}
\newcommand{\nv}{N\,{\sc v}}
\newcommand{\nvl}{N\,{\sc v}\,$\lambda$1240}
\newcommand{\nvll}{N\,{\sc v}\,$\lambda\lambda$1238,1242}

\newcommand{\oil}{O\,{\sc i}\,$\lambda$1304}
\newcommand{\ha}{H{\sc $\alpha$}}
\newcommand{\hb}{H{\sc $\beta$}}
\newcommand{\hg}{H{\sc $\gamma$}}
\newcommand{\hd}{H{\sc $\delta$}}
\newcommand{\hi}{H\,{\sc i}}
\newcommand{\siiv}{Si\,{\sc iv}}
\newcommand{\oiv}{O\,{\sc iv}]}
\newcommand{\siivoiv}{\siiv+\oiv\,$\lambda$1400}
\newcommand{\civ}{C\,{\sc iv}}
\newcommand{\civl}{C\,{\sc iv}\,$\lambda$1549}
\newcommand{\civll}{C\,{\sc iv}\,$\lambda\lambda$1548,1550}
\newcommand{\siiii}{Si\,{\sc iii}]}
\newcommand{\siiiil}{Si\,{\sc iii}]\,$\lambda$1892}
\newcommand{\ciii}{C\,{\sc iii}]}
\newcommand{\ciiil}{C\,{\sc iii}]\,$\lambda$1909}

\newcommand{\heiil}{He\,{\sc ii}\,$\lambda$1640}
\newcommand{\heiilo}{He\,{\sc ii}\,$\lambda$4686}
\newcommand{\mgii}{Mg\,{\sc ii}}
\newcommand{\mgiil}{Mg\,{\sc ii}\,$\lambda$2798}
\newcommand{\oiii}{[O\,{\sc iii}]}
\newcommand{\oiiil}{[O\,{\sc iii}]\,$\lambda$5007}

\newcommand{\feii}{Fe\,{\sc ii}}

\newcommand{\lciv}{L$_\nu$(1549)}

\newcommand{\kms}{km~s$^{-1}$} 

\newcommand{\nhi}{N$_{\rm HI}$}

\newcommand{\nh}{cm$^{-2}$}

\newcommand{\ltsima}{$\; \buildrel < \over \sim \;$}
\newcommand{\simlt}{\lower.5ex\hbox{\ltsima}} 
\newcommand{\gtsima}{$\; \buildrel > \over \sim \;$}
\newcommand{\simgt}{\lower.5ex\hbox{\gtsima}} 

\slugcomment{Prepared for publication in the Astrophysical Journal}

\shorttitle{QSO Spectral Principal Components}

\shortauthors{Shang et al.}

\begin{document}


\title{The Baldwin Effect and Black Hole Accretion:  A Spectral Principal 
Component Analysis of a Complete QSO Sample}

\author{Zhaohui Shang,\altaffilmark{1} 
Beverley J. Wills,\altaffilmark{1}
Edward L. Robinson,\altaffilmark{1}
D. Wills,\altaffilmark{1}
\\
Ari Laor,\altaffilmark{2}
Bingrong Xie,\altaffilmark{3}
and Juntao Yuan\altaffilmark{1}
}

\altaffiltext{1}{Department of Astronomy, University of Texas at
Austin, Austin, TX 78712. shang@astro.as.utexas.edu,
bev@astro.as.utexas.edu, elr@astro.as.utexas.edu}
\altaffiltext{2}{Department of Physics, Technion-Israel Institute of Technology, Haifa, 32000, Israel.} 
\altaffiltext{3}{Department of Physics \& Astronomy, Rutgers University, Piscataway, NJ 08854.} 

\begin{abstract} 

We have performed a spectral principal component analysis (SPCA) for
an essentially complete sample of 22 low redshift QSOs with spectral
data from \lya\ to \ha.  SPCA yields a set of independent principal
component spectra, each of which represents a set of relationships
among QSO continuum and line properties.  We find three significant
principal components, which account for $\sim$78\% of the total
intrinsic variance.  The first component, carrying $\sim$41\% of the
intrinsic variance, represents Baldwin relationships --
anti-correlations between equivalent width of broad emission lines
and continuum luminosity.  The narrow line core (FWHM
$\sim$2000\,\kms) of the broad emission lines dominate this
component.  The second component, accounting for $\sim$23\% of the
intrinsic variance, represents the variations in UV continuum slope,
which is probably the result of dust reddening, with possible
contributions from starlight.  The third principal component is
directly related to the Boroson \& Green ``Eigenvector 1'' (their
first principal component), clearly showing the anti-correlation
between strengths of optical \feii\
 and \oiiil, and other relationships previously found in the \hb\ --
\oiii\ region.  This third component shows the expected strong
correlation with soft X-ray spectral index.  The widths of \ciiil,
\mgiil, and Balmer emission lines are also involved and clearly
correlated, relating this component to black hole mass or Eddington
accretion ratio.  We find an inverse correlation between the
strengths of the UV and optical \feii\ blends, as suggested by some
photoionization models.  We also find correlations of the strengths
of several low-ionization UV lines with  \feii(opt), and a strong
positive correlation of \civl\ with \oiii\ strength.  The wide
wavelength coverage of our data enable us to see clearly the
relationships between the UV and optical spectra of QSOs.  The
Baldwin effect and Boroson \& Green's Eigenvector 1 relationship are
clearly independent.  We demonstrate how Baldwin relationships can be
derived using our first principal component, virtually eliminating
the scatter caused by the third principal component.  This rekindles
the hope that the Baldwin relationships can be used for cosmological
study.

\end{abstract}


\newpage
\eject

\notetoeditor{In this paper, Fig. 1 should be four full pages, in landscape.
Fig.3 and Fig.11 should be in full pages, in portrait.  All the other 
figures can be reduced to one-column width.}

\section{INTRODUCTION}

QSOs probe the universe at the epoch of galaxy formation.  Their
strong, broad emission lines arise predominantly from photoionization
of high speed gas ($10^3$\,\kms\ to $10^4$\,\kms) at $10^2$ to $10^4$
gravitational radii from a supermassive black hole ($\sim 10^8$
M$_\sun$).  The emission line spectrum therefore promises to reveal
the accretion mechanism, the origin of the fuel, and links with the
host galaxy evolution.  In addition, Baldwin (1977) discovered a
relationship between line equivalent width and luminosity, suggesting
that the luminosity may be determined directly from the spectrum,
hence providing an important test of cosmological models at high
redshift, $z$.  Forty years after QSOs' discovery, these promises
have yet to be fulfilled.  There are two main, related, reasons for
this.  (i) Even in the nearby lower-luminosity AGN, the broad line
region (BLR) remains unresolved and its structure can be inferred
only indirectly.  For example, reverberation mapping has shown that
light-travel times from the central continuum source increase with
decreasing ionization (from a few light days, to light months, e.g.,
Netzer \& Peterson 1997).  Emission line strengths and the fact that
absorption along the line-of-sight to the continuum source occurs
only sometimes, show that only a few percent of the center is covered
by BLR gas, and the global filling factor is very small.  (ii) To
first order, emission-line ratios of all UV-optically selected QSOs
are surprisingly similar.   \citet{Bal95} successfully reproduced the
average QSO spectrum -- showing that their similarity was the result
of powerful selection.  Given a BLR with a wide range of gas density,
column density and ionizing flux, different emission lines are formed
predominantly in gas with different physical conditions optimal for
their formation. Thus the integrated line properties of the BLR
spectrum are quite insensitive to the density, ionization, and column
density of BLR gas.  This is true of any BLR model that exhibits a
range of density at each radius; thus the integrated properties of
the BLR cannot clearly distinguish among multi-cloud models
\citep{Bal95, Bot01}, disk-wind models \citep{MC97}, stellar wind, or
bloated stellar atmosphere models (e.g., Scoville \& Norman 1988;
Alexander \& Netzer 1994).

In important details, however, spectra of QSOs are not all similar.
Rather than comparing models with average line profiles and line
ratios, we can learn more from direct observations of the dependence
of QSO spectra on observed parameters that may be related directly to
the central engine.  These dependences, including velocity
information across the broad line profiles, provide a powerful tool
that has yet to be fully exploited.  Two striking sets of
relationships have been found, relating QSOs' optical -- UV emission
lines and soft X-ray continua with fundamental properties of the
central engine.  One is the Baldwin relationship between equivalent
widths of emission lines from the BLR, and the luminous power output,
L.  The other set is the ``Principal Component 1'' (also called
``Eigenvector 1'') relationships, which appear to be related to the
Eddington ratio.  These relationships may allow us to explore the
accretion history of QSOs and, ultimately, the formation of galaxy
spheroids.

The Baldwin effect is an inverse relationship between the equivalent
widths of QSOs' broad emission lines EW, and the UV continuum
luminosity L, EW $\propto$ L$^{\beta}$, spanning 7 orders of
magnitude in luminosity.  It was originally derived using the
\civl\ line \citep{Bal77,Bal89}, and the continuum luminosity at
1549\,\AA.  Other lines show Baldwin relationships (e.g., Kinney,
Rivolo, \& Koratkar 1990, Espey \& Andreadis 1999), with $\beta$ =
$-0.05$ to $-0.3$, the slope apparently increasing with increasing
ionization potential.  The ``line cores'' (emission from low velocity
gas) contribute more than the line wings (Francis et al. 1992 --
hereinafter, FHFC; Osmer, Porter, \& Green 1994; Francis \& Koratkar
1995; Brotherton \& Francis 1999).  It has been suggested that the
Baldwin effect results from a luminosity-dependent spectral energy
distribution, ionization parameter, covering factor, or is the result
of inclination \citep{Pet97, Kor99, FBqc}.  The original hope was
that QSO luminosity could be determined from line equivalent widths,
and that the resulting L -- $z$ relationship could be used to
discriminate among cosmological models.  However, so far, observed
Baldwin relationships show too much scatter.

In a Principal Component Analysis (PCA) of direct, integrated, line
measurements in the \hb\ region of 87 low $z$ QSOs, Boroson \& Green
(1992, hereinafter, BG92) discovered that most of the
spectrum-to-spectrum variance, represented by the first principal
component (Eigenvector 1, hereinafter, BG\,PC1), links a number of
variables:  decreasing broad \hb\ line width corresponds to stronger
\feii\ optical emission (\feii(opt)), weaker \oiiil\ emission, and
increasing \hb\ asymmetry  from stronger red to stronger blue wings.
BG92 suggested that the \feii(opt) -- \oiii\ anticorrelation arose
from an increase in covering factor of dense \feii-emitting gas,
resulting in an increase in the Eddington  accretion
ratio.  The case for higher Eddington ratios was especially supported
by the Laor et al.  (1994, 1997, hereinafter, L94, L97) inclusion of
X-ray spectral index $\alpha_x$ in this principal component: steeper
$\alpha_x$ corresponds to narrower \hb\, etc.  In a virial
interpretation of \hb\ width, narrower \hb\ would correspond to
smaller black-hole mass, hence higher Eddington ratio; steeper
$\alpha_x$ corresponds to higher Eddington ratios as inferred for
Galactic black hole candidates \citep{PDO95}.  Also using a PCA on
integrated line measurements, \citet{Wil99a} showed that these
relationships extend to the UV:  stronger \feii(opt) corresponds to a
larger ratio \siiiil/\ciiil\ indicating higher densities, also weaker
\civl, stronger \nvl, stronger \siivoiv\ feature, and other
properties.  The support for a virial interpretation of \hb\ line
widths has further strengthened the case for an Eddington ratio
interpretation of these BG\,PC1 relationships (Boroson 2002,
hereinafter, B02)

The discovery of supermassive black holes in nearby massive galaxies
\citep{Kor00} and the proportionality between M$_{\rm bh}$ and
spheroid masses \citep{FerMer00,Geb00} has allowed the calibration of
virial masses in nearby QSOs and active galaxies
\citep{Laor98,Geb00}.  Thus, the velocity width of the broad
\hb\ emission line, and a nuclear distance for the \hb-emitting gas
derived from reverberation mapping or photoionization modeling
\citep{Wan99,Kas00}, allow the measurement of M$_{\rm bh}$, hence the
calculation of the maximum (Eddington) accretion luminosity, L$_{\rm
Edd}$, and the Eddington accretion ratio, L/L$_{\rm Edd}$.
Comparison of the space density of local black holes with that of
QSOs at the peak of their luminous output near $z \sim$ 2 -- 3, shows
that nearby galaxies harbor the black-hole relics of QSO activity
(e.g., Marconi \& Salvati 2002; Fabian 1999; Gilli, Salvati, \&
Hasinger 2001).  Together with the M$_{\rm bh}$ -- spheroid mass
relationship this then points to an interrelated evolutionary history
for QSOs and massive galaxies (e.g., Fabian 1999; Silk \& Rees
1998).  Thus an understanding of the above Principal Component
relationships is important for understanding galaxy evolution.

In this paper we extend the above analyses, carrying out a spectral
Principal Component Analysis (SPCA) on the essentially complete
sample of optical-UV spectra of 22 QSOs, the
same sample that we used for the PCA on integrated line measurements
described above.

We briefly describe the sample, observations and reductions in
\S\ref{Obs}.  In \S\ref{Spca} we describe the technique of spectral
PCA.  The results of the SPCA are presented in \S\ref{Results}.  We
first reproduce the main results of the BG92 PCA to illustrate the
power of SPCA.  The principal component spectra covering the \lya\ to
\ha\ region give a striking visual impression of the relationships
among optical and UV emission lines of different ionization stage and
critical density.  Such broad band coverage is essential for
investigating relationships involving broad-band features such as the
entire optical-UV spectral energy distribution, the Balmer continuum
and the blended \feii(UV) emission (``small blue bump'').  
We include some SPCA
simulations to illustrate the non-linear effects of line profile
relationships.  In \S\ref{Disc} we discuss the distinctness and
separation of the emission-line spectral components, and the
interpretation of the ``intermediate line region'' and ``very broad line
region'' as presented by Francis, Brotherton, and collaborators.  We
summarize our results in \S\ref{Summ}, mentioning some implications
for future research, in particular, the usefulness of SPCA for
reducing scatter in Baldwin relationships, and in tracking the
evolution of Eddington ratios in QSOs.  Complementary analyses for
our QSO sample, including results of PCA on direct, integrated,
parameters, have been presented by \citet{Wil99a}, \citet{FWqc},
Wills et al. (1999b,c), \& \citet{Wil00}.

\section{OBSERVATIONS AND DATA REDUCTION \label{Obs}}

We have observed optical-UV spectra for 22 of the 23 PG QSOs making
up the complete optically selected sample investigated by L94 and
L97.  Laor et al. selected all QSOs from the Bright Quasar Survey (BQS)
\citep{Sch83} with redshift $z < 0.4$ and a Galactic hydrogen column
density \nhi $< 1.9 \times 10^{20}$ cm$^{-2}$.  
This sample should be representative of low-redshift, optically selected QSOs,
but subject to the same incompleteness as the BQS \citep{WamPon85,Gol92,Mic01}.
The QSOs in the sample are listed in Table 1 along
with their redshifts, magnitudes, and other parameters.
The low redshift
ensures observational access to the soft X-rays nearest the
wavelength of the ionizing continuum.  The low Galactic absorption
and brightness of this low redshift sample enable accurate
determination of the intrinsic soft X-ray and UV spectra.  In
addition the low redshift and UV coverage allows investigation of the
\lya\ region, with much reduced confusion from intergalactic
absorption lines.  

For the ultraviolet region, we obtained {\em Hubble Space Telescope}
(HST) Faint Object Spectrograph (FOS) spectrophotometry for 16 QSOs
and used archival FOS data for 6 QSOs, covering wavelengths from
below \lya\ to beyond the atmospheric cutoff near 3200\,\AA\ in the
observed frame\footnote{ We were not able to obtain the UV spectrum
of PG1543+489 between 1648\,\AA\ and 2446\,\AA\ (rest frame).  Also
its optical spectral data miss part of the \ha\ red wing.}.
Instrumental resolution is equivalent to $\sim$230\,\kms\ (FWHM).
The FOS pipeline calibration is described at
\url{http://www.stecf.org/poa/FOS/index.html}.

Optical data were obtained at McDonald Observatory, generally with
the Large Cassegrain Spectrograph on the Harlan J. Smith 2.7m
telescope, supplemented by some from the HST FOS.  We were usually
able to get quasi-simultaneous optical observations,  within  a month
of the new HST observations,  to reduce the uncertainty caused by QSO
intrinsic variability.  Observations were made through narrow
(1\arcsec -- 2\arcsec) and wide (8\arcsec) slits.  Standard stars
were observed several times each night.  They were chosen from the
spectral standard stars for HST calibration \citep{Boh96,Boh00} to
ensure consistent calibration between UV and optical data.  Usually
2--3 spectra for each object are required to cover wavelength ranges
from below 3200\,\AA\ to beyond \ha.  Instrumental resolution for the
optical spectra is $\sim$7\,\AA\ FWHM, equivalent to 450\,\kms\ to
$\sim$250\,\kms\ (FWHM) in the \hb\ to \ha\ region.

Higher resolution spectra ($\sim180$\,\kms\ FWHM) of the
\feii(opt)--\hb--\oiii\ region were obtained with the same
spectrograph and telescope at McDonald Observatory and are used only to
determine the redshifts of most of the sample QSOs with
sufficiently strong \oiii\ lines.

Standard packages in IRAF were used to reduce the data.
Wavelength calibration was done by using comparison spectra.  For some
spectra, it was necessary to shift the wavelength scale to match
the wavelengths of night sky lines.
Spectrophotometric calibration was achieved by:  1) using standard
star spectra to calibrate QSO spectra to an absolute flux-density scale,
 2) scaling the narrow slit QSO spectra to match the shape and
absolute flux-density level of the wide slit QSO spectra.  The
photometric accuracy is $\sim$5\%.

\begin{deluxetable}{rlrrccrrr}
\tablecolumns{9}
\tablewidth{0pc}
\tablecaption{The Sample \label{sample}}
\tablehead{
\colhead{Object} & \colhead{$z$\tablenotemark{a}}   & \colhead{m$_B$\tablenotemark{b}}    & \colhead{$\alpha_x$\tablenotemark{c}} & 
\colhead{FWHM(H$\beta$)\tablenotemark{c}}    & \colhead{L$_{\nu}(1549)$\tablenotemark{d}}   & \colhead{W$_{\rm SPC1}$\tablenotemark{e}}    & 
\colhead{W$_{\rm SPC2}$\tablenotemark{e}} & \colhead{W$_{\rm SPC3}$\tablenotemark{e}}\\
\colhead{} & \colhead{}   & \colhead{}    & \colhead{} & 
\colhead{(\kms)}    & \colhead{\small (erg s$^{-1}$ Hz$^{-1}$)}   & \colhead{}    & \colhead{} & \colhead{}
}
\startdata 
PG0947+396&	0.2056	&	16.40	&	$-$1.510	&	4830	&	30.33	&	3.67	&	$-$2.73	&	$-$1.65\\
PG0953+414&	0.2341	&	15.05	&	$-$1.570	&	3130	&	30.59	&	7.31	&	$-$0.99	&	$-$0.88\\
PG1001+054&	0.1603\tablenotemark{f}	&16.13	&$-$2.800	&	1740	&	29.73	&	\nodata	&	\nodata	&	\nodata \\
PG1114+445&	0.1440	&	16.05	&	$-$0.880	&	4570	&	29.80	&	\nodata	&	\nodata	&	\nodata \\
PG1115+407&	0.1541	&	16.02	&	$-$1.890	&	1720	&	30.08	&	$-$2.37	&	$-$6.06	&	2.46\\
PG1116+215&	0.1759	&	15.17	&	$-$1.730	&	2920	&	30.67	&	$-$2.21	&	$-$2.55	&	0.66\\
PG1202+281&	0.1651	&	15.02	&	$-$1.220	&	5050	&	29.51	&	\nodata	&	\nodata	&	\nodata \\
PG1216+069&	0.3319	&	15.68	&	$-$1.360	&	5190	&	30.69	&	$-$0.22	&	7.96	&	$-$2.26\\
PG1226+023&	0.1575	&	12.86	&	$-$0.942	&	3520	&	31.37	&	$-$8.39	&	$-$1.79	&	2.19\\
PG1309+355&	0.1823	&	15.45	&	$-$1.510	&	2940	&	30.21	&	$-$10.15	&	7.36	&	$-$0.67\\
PG1322+659&	0.1675	&	15.86	&	$-$1.690	&	2790	&	30.02	&	6.48	&	3.63	&	$-$0.27\\
PG1352+183&	0.1510	&	15.71	&	$-$1.524	&	3600	&	30.04	&	2.98	&	$-$4.26	&	$-$0.67\\
PG1402+261&	0.165\tablenotemark{g}	&15.57	&$-$1.930	&	1910	&	30.40	&	$-$1.19	&	$-$4.59	&	3.62\\
PG1411+442&	0.0895	&	14.99	&	$-$1.970	&	2670	&	29.56	&	\nodata	&	\nodata	&	\nodata \\
PG1415+451&	0.1143	&	15.74	&	$-$1.740	&	2620	&	29.72	&	6.80	&	3.92	&	2.16\\
PG1425+267&	0.3637\tablenotemark{f}	&15.67	&$-$0.940	&	9410	&	30.42	&	$-$5.01	&	1.29	&	$-$6.59\\
PG1427+480&	0.2203	&	16.33	&	$-$1.410	&	2540	&	30.20	&	7.94	&	$-$1.50	&	$-$2.84\\
PG1440+356&	0.0773	&	15.00	&	$-$2.080	&	1450	&	29.90	&	7.33	&	6.16	&	7.71\\
PG1444+407&	0.267\tablenotemark{g}	&15.95	&$-$1.910	&	2480	&	30.63	&	$-$9.49	&	1.75	&	2.32\\
PG1512+370&	0.3700\tablenotemark{f}	&15.97	&$-$1.210	&	6810	&	30.66	&	$-$0.02	&	$-$0.03	&	$-$7.01\\
PG1543+489&	0.400\tablenotemark{g}	&16.05	&$-$2.110	&	1560	&	30.75	&	$-$6.27	&	$-$2.96	&	2.83\\
PG1626+554&	0.1317	&	16.17	&	$-$1.940	&	4490	&	30.11	&	2.81	&	$-$4.61	&	$-$1.12\\

\enddata 

\tablenotetext{a}{From measurements of \oiii\ after removing
\feii\ emission in our separate higher resolution spectra unless
noted}

\tablenotetext{b}{From \citet{Sch83}}

\tablenotetext{c}{From \citet{Lao97}. $\alpha_x$ is soft X-ray spectral index.}

\tablenotetext{d}{Continuum luminosity at 1549\AA, from measurements
of our UV spectra.  We use H$_0$ = 50 \kms Mpc$^{-1}$, and q$_0$ = 0.5.}

\tablenotetext{e}{Weights of principal components from the SPCA of
18 spectra covering \lya\ to \ha\ in \S\ref{Results}.}

\tablenotetext{f}{From measurements of \oiii\ in the spectra
of this study.  \feii\ emission was not removed.}

\tablenotetext{g}{From measurements of \hb\ and other emission lines
in the spectra of this study.  These objects have very weak \oiii. }

\end{deluxetable}

UV and optical spectra for each object were then combined in the
observed frame.  Before combination, some spectra were scaled by a
few percent to match, in the overlap region, the continuum level of
the spectrum with the best flux-density calibration, usually the HST
spectra.  Galactic reddening was removed using the \hi\ column
densities \nhi\ from accurate 21~cm data (references in  L94, L97)
and an empirical relationship \nhi $=53 \times 10^{20}$ E(\bv)
\nh\ \citep{Pre95}.  The column densities are low with uncertainties
of $\pm 10^{19}$ cm$^{-2}$,  insuring small
corrections for Galactic absorption, especially in the UV.  
The error in flux density introduced by this correction 
is less than 2\% even at the shortest wavelengths.  The
dominant
uncertainty in flux density arises from the scatter in the \nhi\ --
E(B-V) relationship, and is $\sim$12\% in the worst case, but generally
significantly less.  

After subtraction of optical \feii\ emission blends
with the same \feii\ template and method used by BG92, 
we used the higher resolution spectra to define a
rest-frame wavelength scale referred to a wavelength of
5006.8\,\AA\ for \oiii\ and applied the redshift to the combined
spectra.  Redshifts for other QSOs were obtained from the lower
resolution spectra in this study (Table~\ref{sample}).  We present
our combined spectra in Fig.~\ref{allspectra1}.

\begin{figure}
\epsscale{0.9}
\plotone{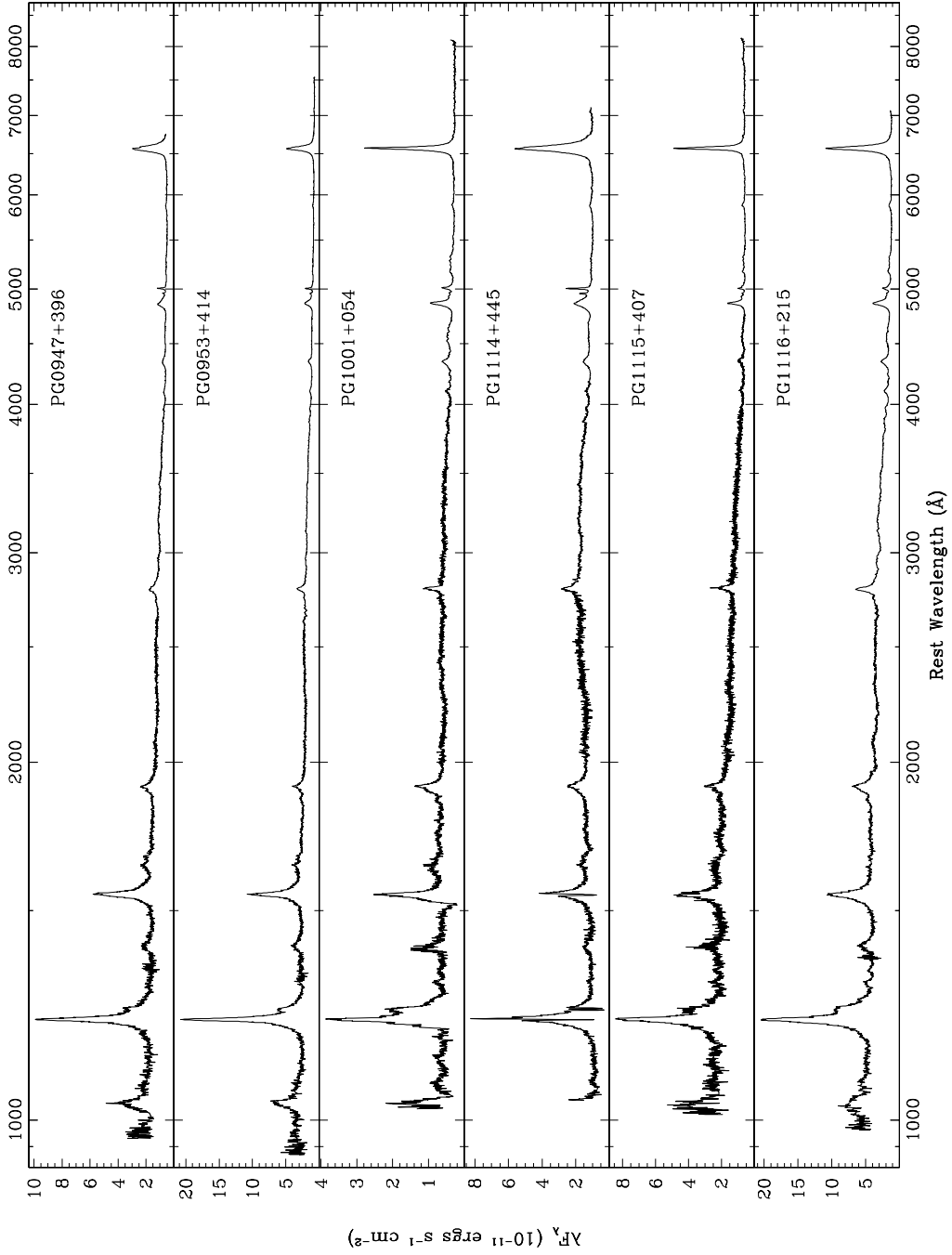}
\caption{Combined rest-frame UV-optical spectra of the QSOs in our sample.
Galactic absorption lines are removed except for some in the far UV.  Galactic
reddening is also removed.
\label{allspectra1}}
\end{figure}

\begin{figure}
\figurenum{1}
\epsscale{.90}
\plotone{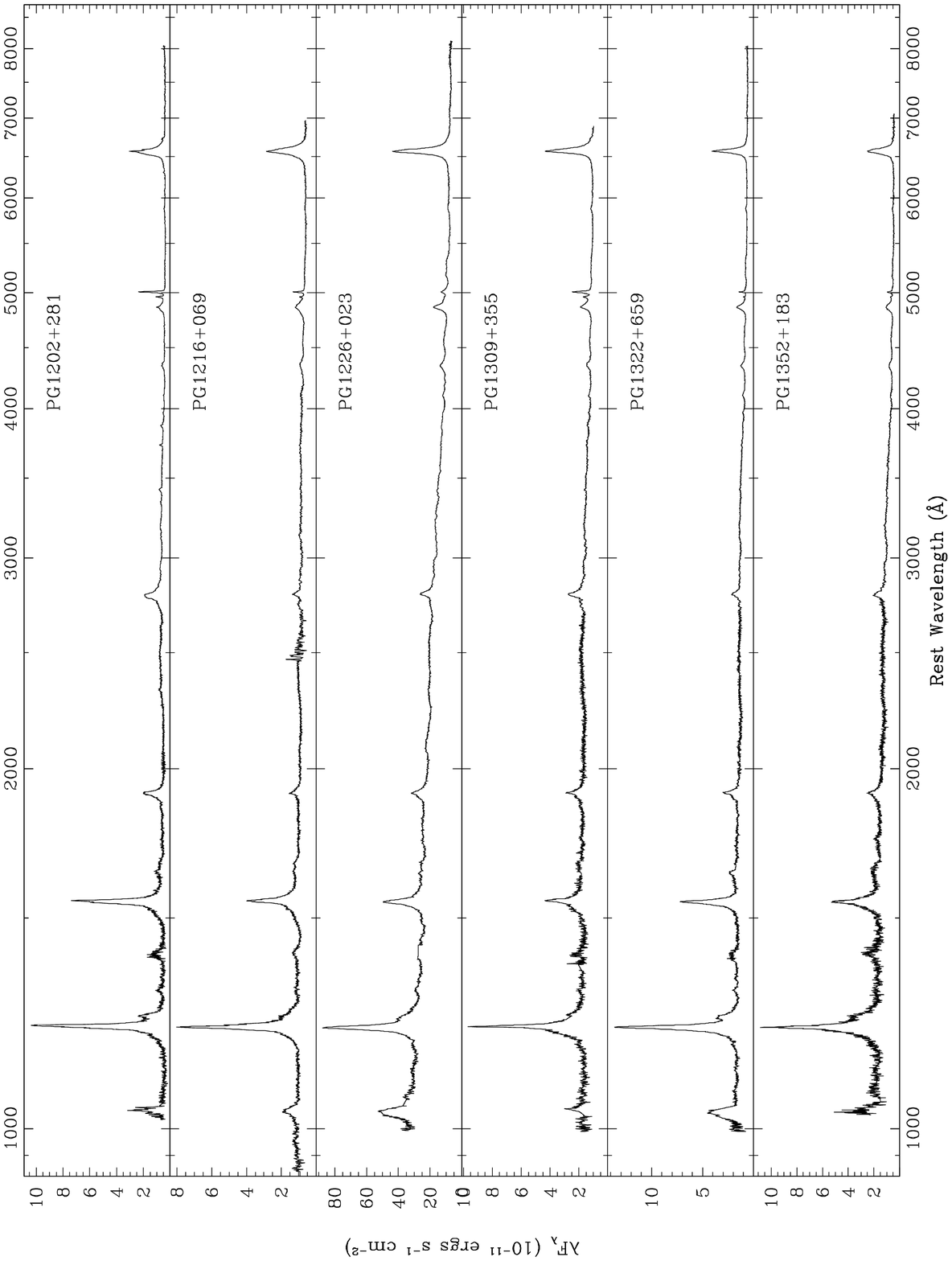}
\caption{Continued
\label{allspectra2}}
\end{figure}

\begin{figure}
\figurenum{1}
\epsscale{.90}
\plotone{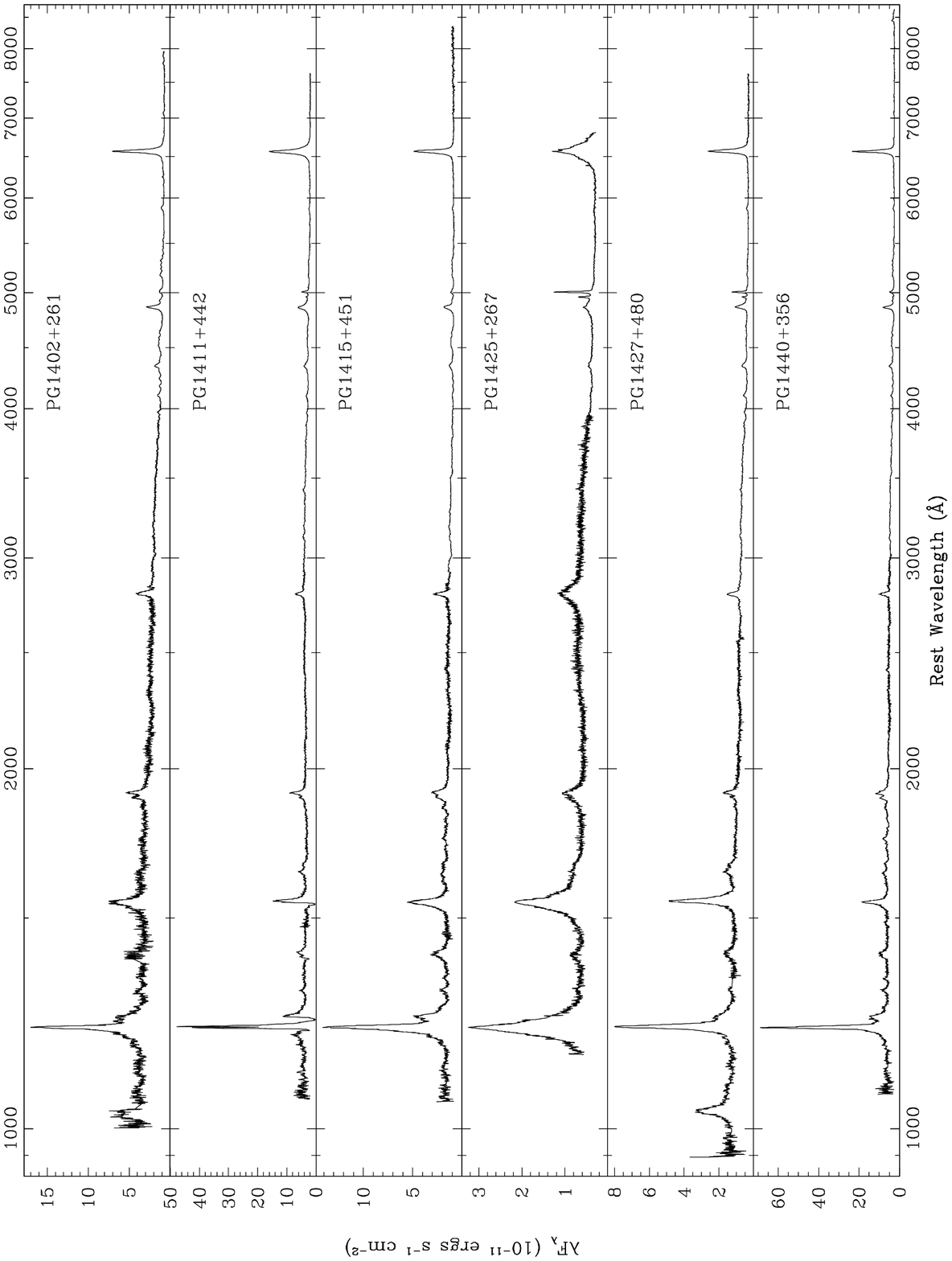}
\caption{Continued
\label{allspectra3}}
\end{figure}

\begin{figure}
\figurenum{1}
\epsscale{.90}
\plotone{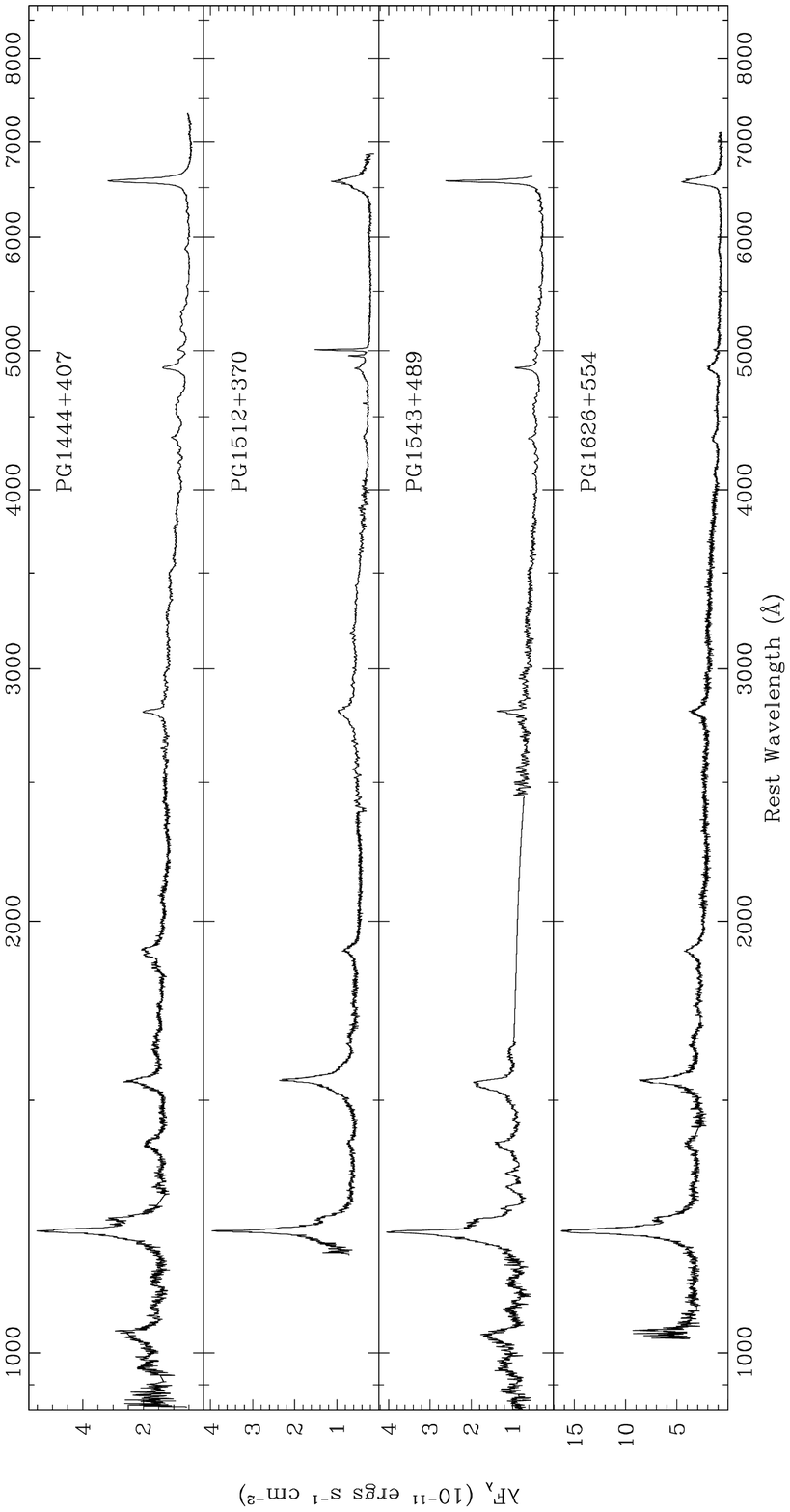}
\caption{Continued
\label{allspectra4}}
\end{figure}

\section{SPECTRAL PRINCIPAL COMPONENT ANALYSIS \label{Spca}} 

Principal component analysis (PCA) is a powerful tool for
multivariate analysis.  Given a sample of $n$ QSOs, each with $p$
measured variables, $x_i$ (e.g., line width, luminosity, soft X-ray
spectral index $\alpha_x$, etc.), PCA defines a set of $p$ new
orthogonal variables PC$_j$.  The PC$_j$ are linear combinations of
the original variables:

$$ {\rm PC}_j = \sum^p_{i=1} a_{ij}x_i \eqno(1)$$

\noindent
ordered according to the fraction of the total sample variance
accounted for by each PC$_j$, e.g., PC$_1$ accounts for more variance
than any other, PC$_2$ accounts for the next largest amount of
variance, etc..  If there are strong linear relationships among the
original variables, each of these relationships will be represented
by a principal component, and fewer principal components will be
required to describe the total variation among QSOs, thus providing a
simpler description of the dataset.  If the measured variables are
unrelated, there will still be $p$ PC$_j$s; but no simplification will
result; nothing will have been accomplished.  The hope is that any
PC$_j$ accounting for a significant fraction of the total sample
variance might be related to one or more underlying fundamental physical
parameters, giving some physical insight into the cause of the
variations (see BG92, B02).  The PCA technique is explained in more
detail and applied to integrated, direct, measurements of the present
sample of 22 spectra, by \citet{FWqc} and \citet{Wil99b}.

In spectral PCA (SPCA), the whole spectrum is divided into $p$ small
wavelength bins, and each input variable $x_i$ is the flux in the
$i$th wavelength bin.  The $j$th principal component from this
analysis can be represented as spectra of the coefficients $a_{ij}$
of each $x_i$ (see \S\ref{Results}).  Features of the same sign in
the $a_{ij}$ spectrum are positively correlated, while those with
opposite sign are anticorrelated.  One of the advantages of SPCA is
that correlations can be investigated without parameterizing the line
profiles or defining the continua.  
In addition, unlike the ``composite spectrum
analysis'' which shows average QSO properties,
the SPCA analysis keeps the information for individual
QSOs, e.g., the weights of the principal component spectra 
for each QSO (Table~\ref{sample}).
The original (normalized) spectra
in the sample can therefore be reconstructed by adding the weighted
principal component spectra to the mean spectrum.   The SPCA technique is
described in more detail by FHFC and references therein.

We have rebinned the rest-frame spectra in equal logarithmic
wavelength intervals, representing equal intervals in velocity
space.  Individual spectra are first normalized by their mean flux
density; this means that the resultant principal component spectra
represent flux density variations relative to this mean -- for
emission lines the integral over the line in the principal component
spectra represents essentially equivalent width
variations.  Next, the mean and standard deviation spectra are
formed, and the mean spectrum subtracted from each QSO spectrum.
This is equivalent to moving the dataset to the origin of the
multidimensional space represented by the $x_i$.  We used the code
developed by FHFC ({http://www.mso.anu.edu.au/\verb+~+pfrancis/}).
The resulting principal components are ordered according to the
fraction of the total sample variance that they account for.

\section{RESULTS \label{Results}}

\begin{figure}
\epsscale{0.7}
\plotone{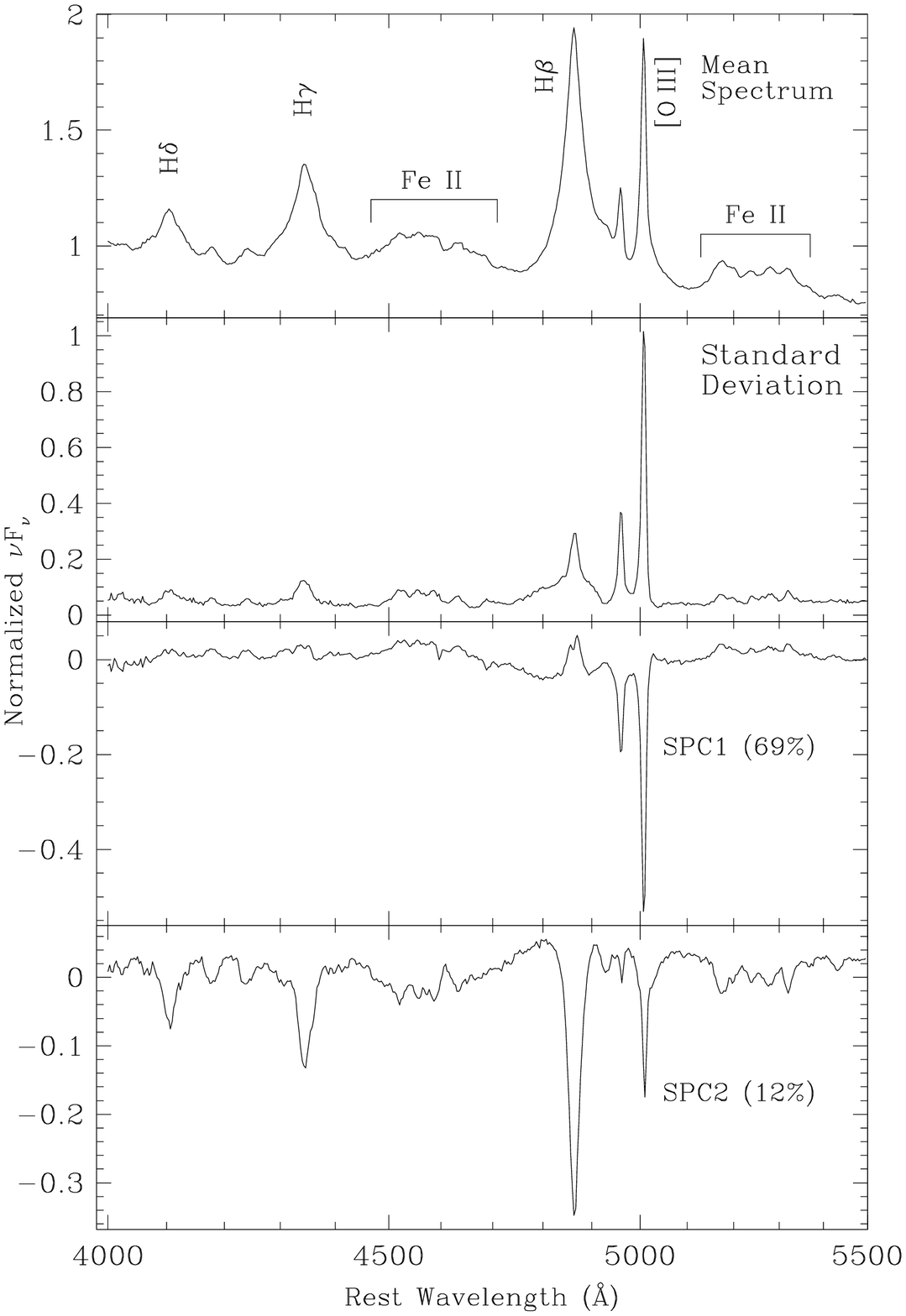}
\caption{SPCA results in the \hb\ region for 22 QSOs.  The
ordinate is $\nu F_\nu$, with each QSO spectrum normalized by its
mean flux density.  Features that extend in the same or opposite
directions are positively or negatively correlated, respectively, in
the SPC spectra.  The first principal component is similar to the
\citet{BG92} first principal component.  The ``W'' shape of \hb\ in
SPC1 indicates the width increases with stronger
\oiii\ (\S\ref{Sim}).  The second principal component shows the
correlations of Balmer lines.  The number in the parentheses
indicates the percentage of the total intrinsic variance that each
principal component accounts for (see \S\ref{Results}).  It shows
that the first principal component is much more important than the
second one in this region.
\label{optspca}}
\end{figure}

We first apply SPCA to our sample in the \hb\ region and compare the
results (Fig.~\ref{optspca}) with the well known BG\,PC1.  As
expected, the strong anticorrelation between \oiiil\ and
\feii\ appears in the first principal component, indicated by the
oppositely directed \oiii\ and \feii\ features.  The ``W'' shape of
\hb\ in the first principal component indicates that the width of
\hb\ increases with increasing \oiii\  equivalent width (see
\S\ref{Sim} for simulations).  The second principal component shows
that the equivalent widths of the Balmer lines are correlated with
each other and these lines are narrower than in the mean spectrum.
Neither principal component is exactly equivalent to
BG92's principal components 1 and 2 because they used integrated line
measurements and other parameters in their PCA analysis, such as
absolute magnitude M$_{\rm v}$ and optical--X-ray spectral index
$\alpha_{ox}$.

\begin{figure}
\epsscale{.95}
\plotone{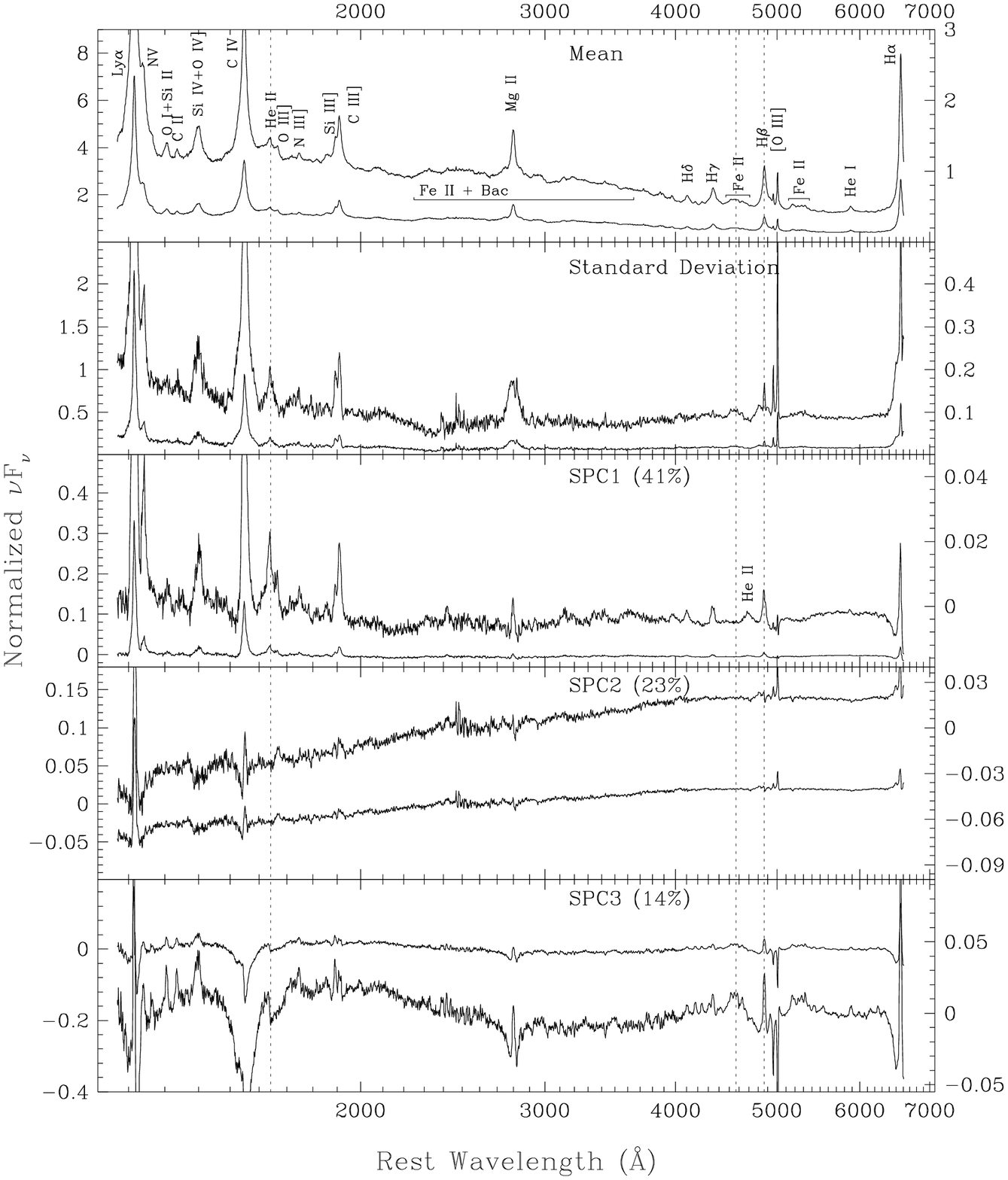}
\vspace{-15mm}
\caption{SPCA results for the UV-optical spectra of 18 QSOs.  The
numbers in parentheses are the percentages of the sample intrinsic variance
that are accounted for by each
principal component.  Each panel presents the spectrum with the full
ordinate range (left-hand scale), and scaled to show details of the
weaker features (right-hand scale).  The dashed vertical lines
indicate the alignment at the \heiil\ feature, at the peak of the
\feii(opt) feature, and at \hb. 
\label{spca}}
\end{figure}

We next apply SPCA to the sample over the wavelength range from
\lya\ to \ha.  The three absorption-line  QSOs (PG\,1001+054,
PG\,1114+445 and PG\,1411+442) and PG1202+281, which is an extreme
object in the sample, are excluded from this analysis with broad
wavelength coverage (for details, see  \S\ref{Bal}).
The results using 18 objects\footnote{For PG\,1543+489, we linearly
interpolate the spectrum over the missing data from 1648\,\AA\ to
2446\,\AA.  \ SPCA shows essentially the same results whether we
include or exclude this object.} are presented here.  The principal
component spectra are shown in Fig.~\ref{spca}, where they are
compared with the mean and standard-deviation spectra.

Compared with the mean spectrum, the standard deviation spectrum has
stronger emission from low velocity gas. \hb\ shows a narrow
component and a broad component with a stronger blue wing.  The
\lyal\ and \nvl\ lines are more clearly resolved than in the mean
spectrum; similarly the \siiiil\ and \ciiil\ emission lines.
\mgiil\ appears not to follow this pattern.  \feii\ emission near
\hb\ shows some sharper features in the standard deviation spectrum
than in the mean spectrum, indicating that \feii\ emission also has
different kinematic components.  That the line cores contribute most
to spectrum-to-spectrum variance has been noted before (e.g., FHFC;
Brotherton et al.\ 1994a).  
The concave shape of the overall standard deviation spectrum arises
simply from the dispersion of continuum slopes among individual 
spectra, because each is normalized by its mean flux density.

The total variance of the sample is the sum of the intrinsic variance
and the variance from the observational and instrumental noise in the
original spectra.  We are interested in the intrinsic variance.  We
first estimate the contribution from noise, for each QSO spectrum.
We create a noise spectrum by shifting the original spectrum by two
pixels, form the difference spectrum of the two, remove sharp spikes
around strong line features, and divide by $\sqrt{2}$.  We then
calculate the noise contribution to the variance by summing the
squares of the fluxes of each bin in the noise spectrum.  We then
calculate the total variance for each QSO spectrum, by summing the
squares of the differences between that spectrum and the mean
spectrum.  Similarly we calculate the variance between each observed
spectrum, and its reconstruction derived from the mean plus the
principal component spectra appropriately weighted for that QSO.
After removing the noise contribution, we calculate the fraction of
the total intrinsic variance accounted for by the addition of each
subsequent principal component, as described by FHFC.  The results
are summarized for the whole spectrum, the UV, and the 2100\,\AA\ --
6607.5\,\AA\ region, in Table~\ref{pcfrac}.  We find that the first,
second, and third components account for 41\%, 23\%, and 14\% of the
total intrinsic variance, for a total of 78\% of the intrinsic
variance.


\begin{table}
\begin{center}
\caption{Fractional Component Contributions to the Total Intrinsic Variance 
\label{pcfrac}}
\begin{tabular}{lccc}
\tableline\tableline
Component(s)  & \multicolumn{3}{c}{ Wavelength range (\AA)}  \\
                & 1171--6607.5 & 1171--2100 & 2100--6607.5      \\
\tableline
SPC1      &     0.41    &    0.49    &    0.08 \\
SPC2      &     0.23    &    0.18    &    0.44\\
SPC3      &     0.14    &    0.13    &    0.20\\
SPC3-10   &     0.31    &    0.29    &    0.41    \\

\tableline

\end{tabular}

\end{center}
\end{table}

\subsection{The First Principal Component: Emission-Line Cores \label{Spc1}}

\begin{figure}
\epsscale{0.4}
\plotone{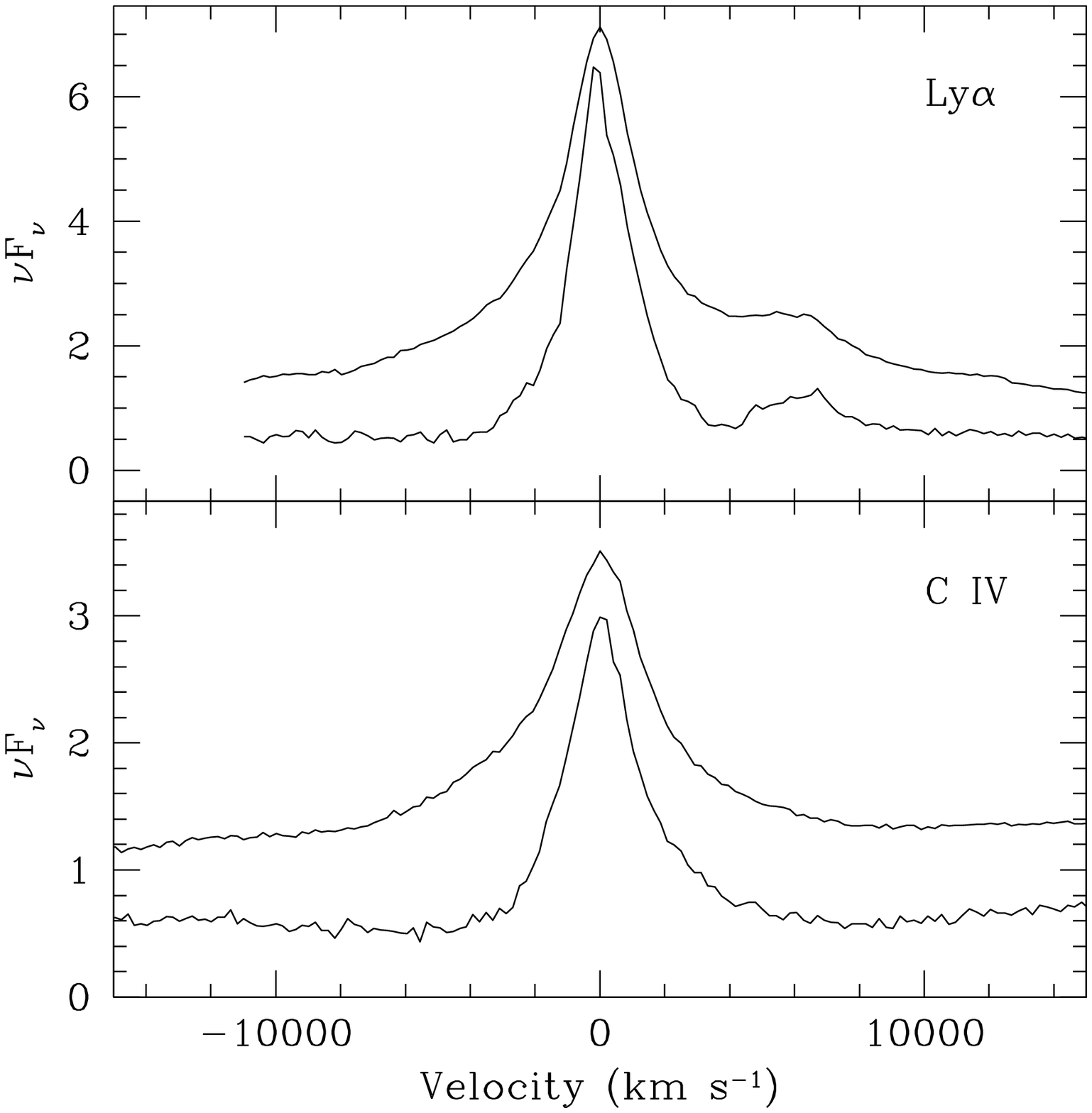}
\caption{Comparison of SPC1 line profiles (lower spectra) with line
profiles of the mean spectrum, showing the blueshifted bases in
\lya\ and \civ.
\label{blueshift}}
\end{figure}

The most striking feature of the first principal component spectrum
(Fig.~\ref{spca}, SPC1) is the correlation of the flux from gas of
relatively low radial velocity  (hereinafter ``line-core
component'').  The widths (FWHM) of the emission line features in
this component are given in Table~\ref{fwhm}, and are typically
2000\,\kms\ when corrected for blended doublets.  By contrast
\heiilo\ is very broad, with FWHM $\approx 4900$\,\kms.  The peaks of
these features are at the systemic redshift and are symmetric.
Relative to these narrow profiles, the broad line profiles of the
mean spectrum appear to have a blue-shifted ``base'', at least for
\lya\ and \civ\ (Fig.~\ref{blueshift}) as noted in higher-redshift
spectra by FHFC and \citet{Bro94a}.

\begin{table}
\begin{center}
\caption{SPC1 and Mean Spectrum Line Widths
\label{fwhm}}
\begin{tabular}{lccc}
\tableline\tableline
		&	\multicolumn{2}{c}{SPC1} & \multicolumn{1}{c}{Mean} \\
		 	\cline{2-3}
Line            &       FWHM    &  corr. FWHM\tablenotemark{a} & FWHM\\
\tableline
\lya            &       1940  & $\cdots$	& 2910\\
\nvll           &       3150  & 2911 		& $\cdots$\\  
\civll          &       2390  & 2313		& 3680\\
\ciiil          &       2390  & $\cdots$	& 4480\\
\mgii\,$\lambda\lambda$2796,2803          &       \phm{:}1370:  & \phm{:9}970:	& 2860\\
\hd             &       \phm{:}2370:  & $\cdots$ 	& 2760\\
\hg             &       2250   & $\cdots$	& 3400\\
\heiilo         &       4880   & $\cdots$    	& $\cdots$\\
\hb             &       1680   & $\cdots$	& 2600\\
\ha             &       1440:  & $\cdots$ 	& 2300\\

\tableline
\end{tabular}

\tablenotetext{}{Note -- FWHMs (\kms) are estimated by directly measuring
the widths between the half maxima of line profiles.  \lya\ and \nv\
are not deblended in the mean spectrum, and \heiil\ cannot be measured there.}
\tablenotetext{a}{Corrected FWHM assume a 1:1 doublet intensity ratio
for \nv\ and \civ, 2:1 for \mgii.}

\end{center}
\end{table}

The strength of broad \heiilo\ is correlated with that of the line
cores (but absent in SPC3).  The revealing of \heiilo\ in SPC1 is
striking, because it is not obviously present in either the mean or
standard deviation spectrum.  We cannot exclude a broad \heiil\ line
corresponding to \heiilo\ as it is blended with narrower \heiil\ and
other emission lines.  Emission from the broad \feii\ UV and optical
blends appears weak or absent.

In the UV, the line-core spectrum appears to be essentially the same
as the first principal component of the SPCA analysis of 232 UV
spectra of the Large Bright QSO Survey (LBQS) by FHFC.  \citet{Fra95}
followed up and confirmed the FHFC suggestion that this emission may
be responsible for the Baldwin effect, by deriving the line-core
equivalent width dependence on luminosity.  This was 
demonstrated more qualitatively by \citet{Osm94},
and \citet{Gre98} also suggested this for \lya\ and \civ.  
However these published studies cover only the strong UV lines between
$\sim$1150\,\AA\ and 2000\,\AA.  

\begin{figure}
\epsscale{1} 
\plottwo{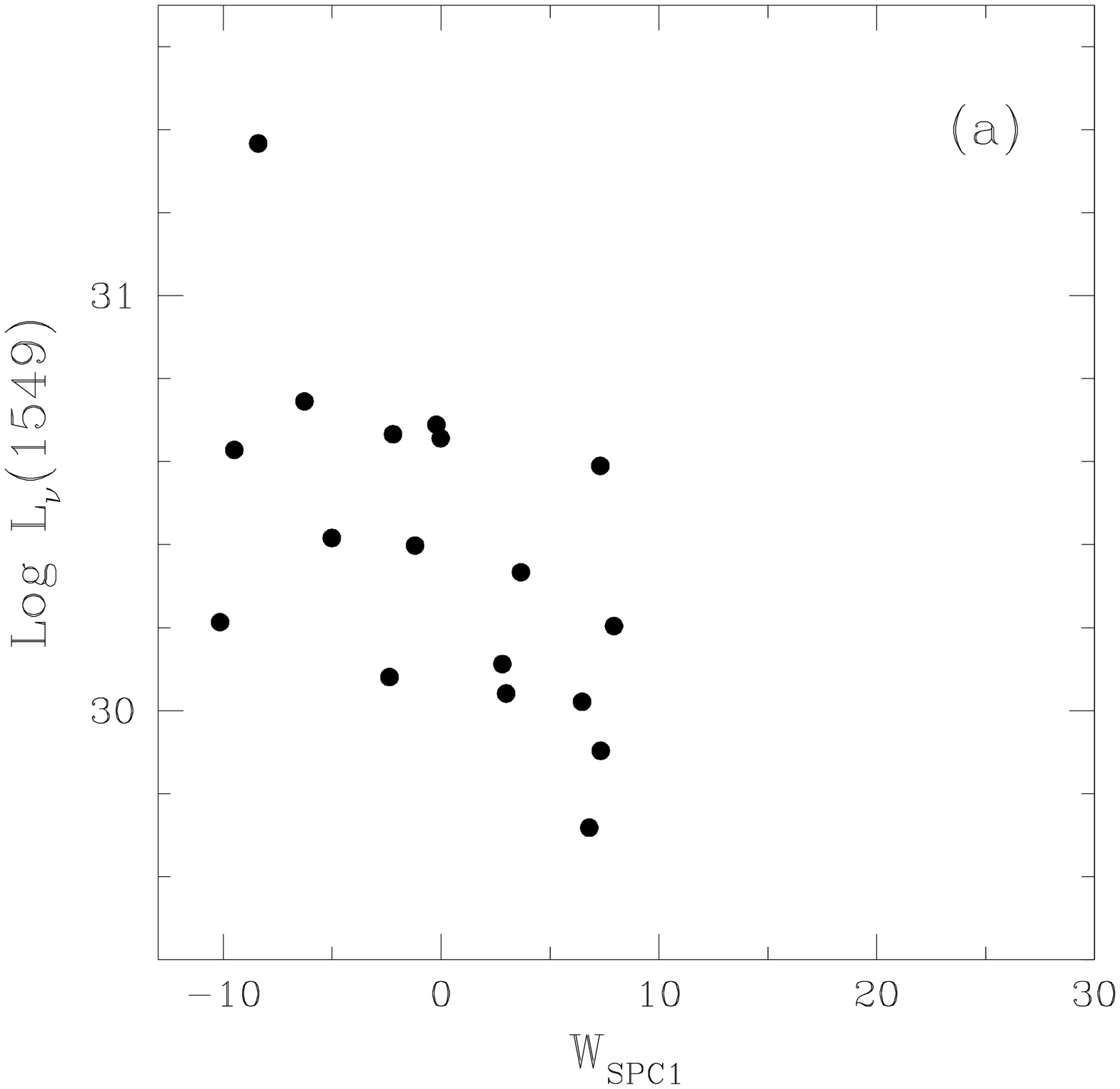}{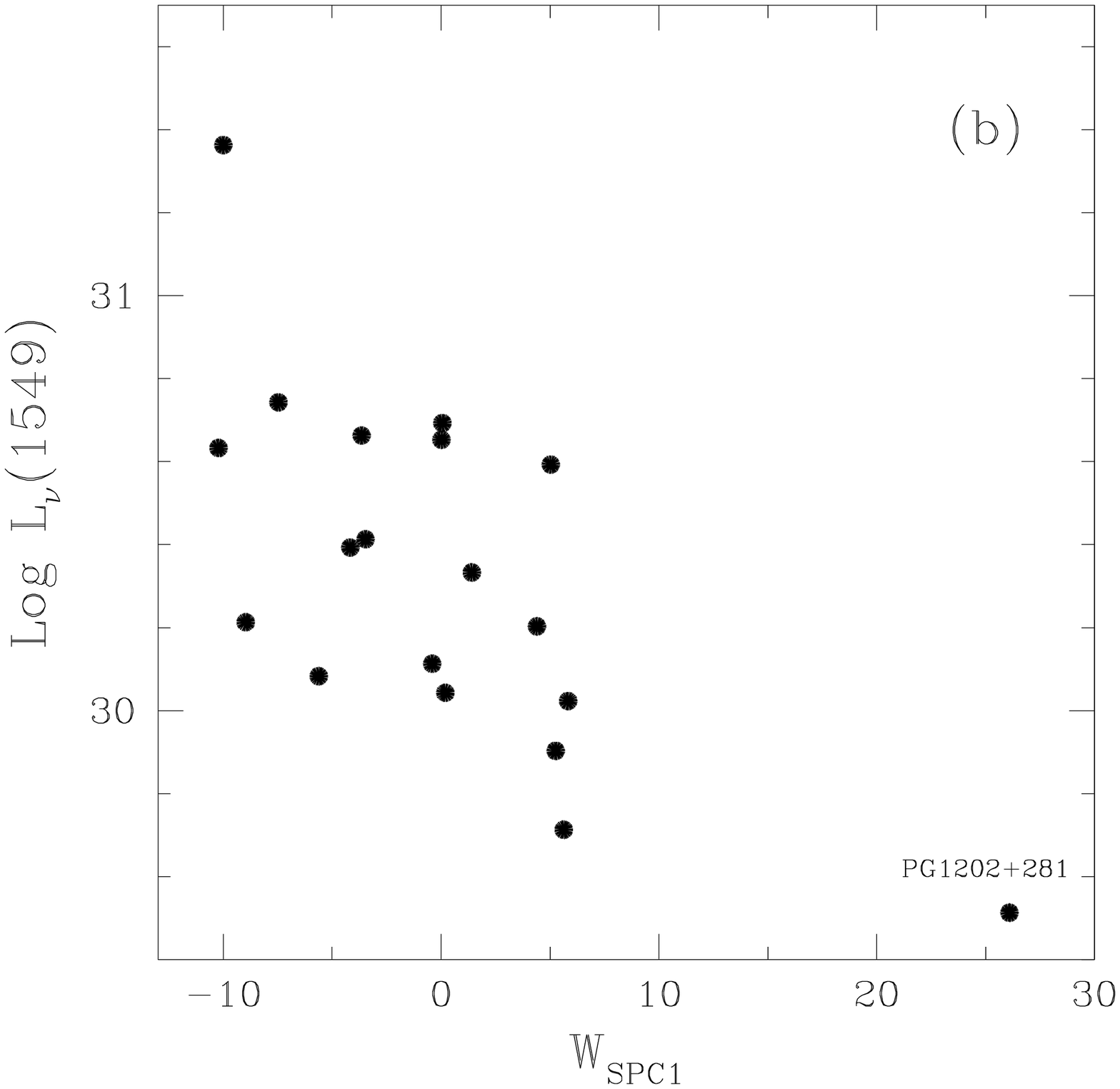}

\caption{Continuum luminosity at \civ\ vs.  weight of SPC1 (line-core
component) from {\em separate} SPCA analyses  (a) for 18 QSOs; (b) including
PG\,1202+281, the lowest luminosity QSO in our sample, included.
Both correlations indicate a Baldwin Effect.  The Pearson
correlation coefficients are 0.58 and 0.69, and the two-tailed
probability of these correlations arising by chance are 1.1\% and
0.13\%, respectively.
\label{spc1-L}}
\end{figure}

Is our SPC1 really the same as that for the LBQS, and does it depend
on luminosity as expected for the Baldwin Effect?  Fig.~\ref{spc1-L}a
and Table~\ref{corr} show that the weight of this line-core component
for each QSO is indeed anticorrelated with (logarithmic)
L$_\nu$(1549), the continuum luminosity at 1549\,\AA.  For 18 QSOs,
the Pearson and Spearman rank (two-tailed) probabilities of this
correlation arising by chance from uncorrelated variables are 1.1\%
and 2.0\%, respectively.  We have redone the SPCA, including the
extreme, low luminosity QSO PG\,1202+281 (Fig.~\ref{spc1-L}b) and
these probabilities become 0.1\% and 0.5\%, respectively.  A Bayesian
analysis gives $<$27\% and $<$7\% probabilities of the correlation arising
by chance for the two cases above.
Since all the spectra are normalized, this is actually an
anticorrelation between line equivalent width and luminosity.
\citet{Wil99a} reported, for the same sample, evidence of the Baldwin
effect in the second principal component of their PCA analysis of
direct line measurements that do not separate the narrow and broad
components (Wills et al. 1999b \&  Francis \& Wills 1999 give
complementary details).  We now see that the correlation they found
was actually a correlation with the equivalent widths of the line-cores.

\begin{table}
\begin{center}
\caption{Correlation Coefficients and Probabilities\tablenotemark{a}
\label{corr}}
\begin{tabular}{lllllllll}
\\[-2pt]
\tableline\tableline
   &W$_{\rm SPC1}$\tablenotemark{b}        &       W$_{\rm SPC2}$\tablenotemark{b}
        &       W$_{\rm SPC3}$\tablenotemark{b}&   $\alpha_x\tablenotemark{c}$
        &  \hb\ FWHM    & L$_{\nu}(1549)$ & M$_{\rm BH}$ & L/L$_{\rm Edd}$      \\
\tableline
W$_{\rm SPC1}$    	&$\cdots$&$\cdots$&$\cdots$&$\cdots$&$\cdots$& $\cdots$  &$\cdots$&$\cdots$\\
W$_{\rm SPC2}$    	&  0     &$\cdots$&$\cdots$&$\cdots$&$\cdots$&$\cdots$&$\cdots$&$\cdots$\\
W$_{\rm SPC3}$		&  0     &    0   &$\cdots$& $\cdots$    &$\cdots$	&$\cdots$& $\cdots$ & $\cdots$ \\
$\alpha_x$&$-$0.167& +0.143 & {\bf $-$0.702}\tablenotemark{d} &$\cdots$&$\cdots$  &$\cdots$& $\cdots$ & $\cdots$ \\
		&	 &	  & {\small (0.12\%)} &   &  &   &   &   \\
\hb\ FWHM	& $-$0.112  & +0.101  & {\bf $-$0.838}  & {\bf +0.731}  &$\cdots$&$\cdots$& $\cdots$      & $\cdots$      \\
		&	 &	  & {\small ($<$0.01\%)} & {\small (0.06\%)}  &  &   &   &   \\
L$_{\nu}(1549)$	& {\bf $-$0.583}  & $-$0.143  & $-$0.151  & +0.433  &+0.200   &$\cdots$&$\cdots$&$\cdots$\\
		& {\small (1.1\%)} &	  &  &  &  &   &   &   \\
M$_{\rm BH}$    & $-$0.246  & +0.137  & {\bf $-$0.711}  & {\bf +0.747}  &{\bf +0.932}   & +0.335  &$\cdots$&$\cdots$\\
		&	 &	  & {\small (0.09\%)} & {\small (0.04\%)} & {\small (0)}  &   &   &   \\
L/L$_{\rm Edd}$ & $-$0.024  & +0.002  & {\bf +0.722}  & {\bf $-$0.499}  &{\bf $-$0.909}   & +0.118  & {\bf $-$0.777}  &$\cdots$\\
		&	 &	  & {\small (0.07\%)} & {\small (3.5\%)} &  {\small (0)}  &   & {\small (0.01\%)}  &   \\

\tableline
\end{tabular}

\tablenotetext{a}{The table gives the Pearson correlation coefficients
for an analysis of 18 QSOs
and the two-tailed probability (in parenthesis) of that correlation arising by chance.
Signs indicate a positive or negative correlation.
}
\tablenotetext{b}{Weights of principal components for each QSO, for the
SPCA of the \lya\ to \ha\ spectra of 18 QSOs.}
\tablenotetext{c}{Soft X-ray spectral index (L97).}
\tablenotetext{d} {3C\,273 (PG\,1226$+$023) has a significant radio-jet 
contribution to $\alpha_x$ and shows significant slope variations (L94).
Excluding this QSO, the Pearson coefficient and probability are
$-$0.894 and 10$^{-6}$.}

\end{center}
\end{table}

\begin{figure}
\epsscale{0.5}
\plotone{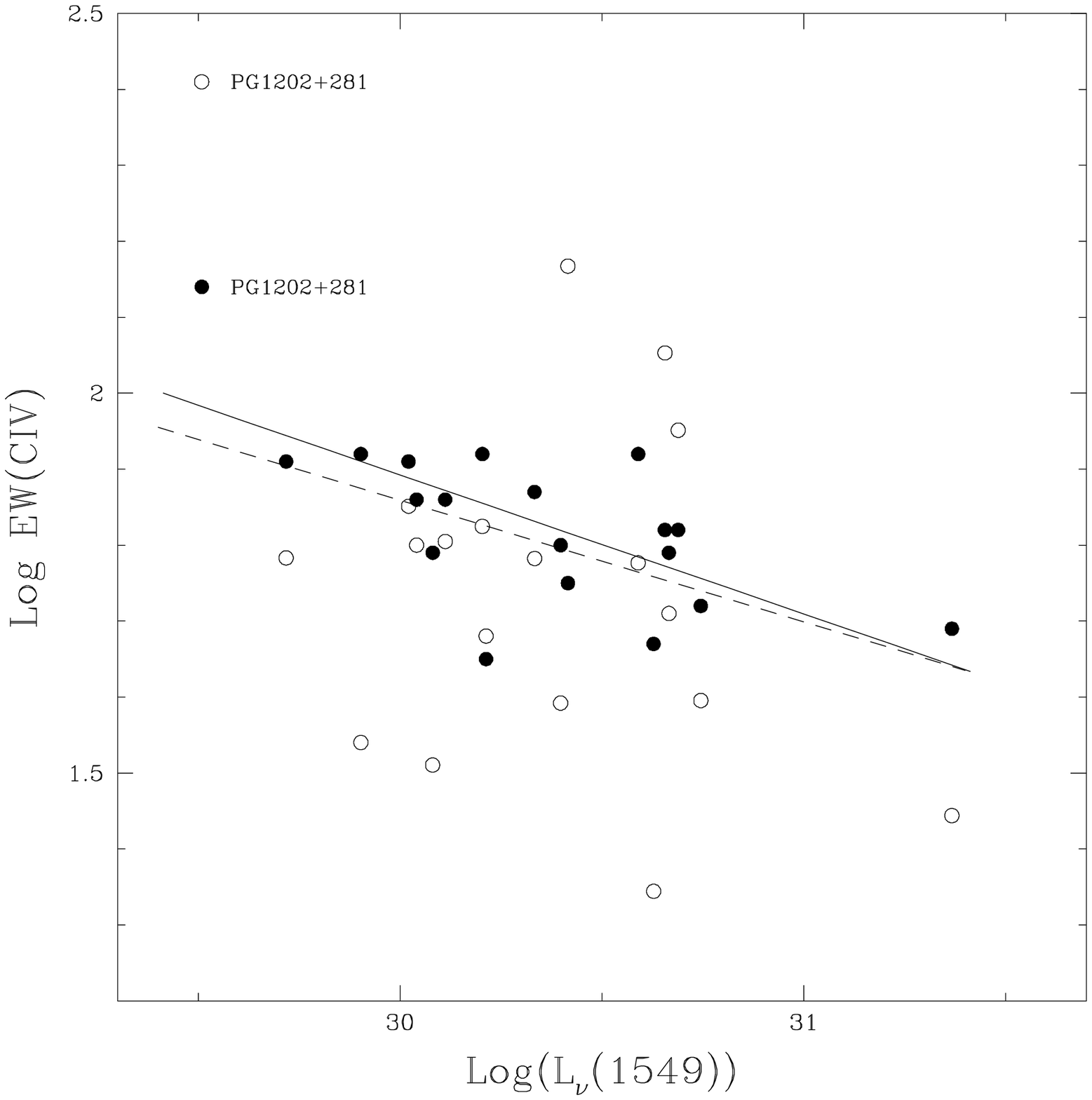}
\caption{Baldwin Relationship for \civ.  The filled circles represent
EWs derived from SPC1 and the mean spectrum (\S\ref{Spc1}).  The
solid line indicates the least-squares regression of the SPC1-derived EW on 
log L$_{\nu}(1549)$.  The open
circles represent the integrated, direct, EW measurements from the original
spectra \citep{FWqc}.  The dashed line is the least-squares fit of
the Baldwin relation from \citet{Kin90}, showing the good agreement
in both slope and normalization.  { (Note: PG\,1202+281 is
excluded from this SPCA analysis.  Its SPC1 weight and therefore
its derived EW in this plot, is estimated based on the tight
relationship between the weight of the first principal component 
in this analysis
and that in the analysis with PG\,1202+281 included
(Fig.~\ref{spc1-L}a,b)).}
\label{EW-L}}
\end{figure}

To test whether our correlation is actually consistent with Baldwin
relationships found in the UV spectra of much larger samples, we
measure the equivalent width of \civ\ from the SPCA results for 19
QSOs.  ${\rm EW}({\mbox \civ}) = ({\rm F}_{\rm mean} + {\rm W}_{\rm
SPC1} {\rm F}_{\rm SPC1})/ ({\rm C}_{\rm mean} + {\rm W}_{\rm SPC1}
{\rm C}_{\rm SPC1})$, where ${\rm F}_{\rm mean}$ and ${\rm C}_{\rm
mean}$ are the \civ\ flux and the flux density of the fitted local
continuum in the mean spectrum, ${\rm F}_{\rm SPC1}$ and ${\rm
C}_{\rm SPC1}$ are those in the first principal component spectrum,
and  ${\rm W}_{\rm SPC1}$ is the weight of SPC1 for each QSO,
evaluated from SPCA.  We deblend the \civ\ region to exclude
\heiil\ and {O\,{\sc iii}]\,$\lambda$1664} in the mean spectrum, and
\civ\ is narrow enough to be separated from \heiil\ and {O\,{\sc
iii}]\,$\lambda$1664} in SPC1.  This procedure is equivalent to
measuring the spectra reconstructed with only the mean spectrum and
SPC1, and therefore includes the equivalent width variation of the
line cores, but excludes scatter that would otherwise be introduced
by orthogonal principal components.  The filled circles in
Fig.~\ref{EW-L} show the anticorrelation between this EW(\civ) and
L$_\nu$(1549).  The slope of the least-squares fit is $-0.18\pm0.05$,
which agrees with $-0.16\pm0.06$ obtained by \citet{Kin90}.  This
could be coincidence for our small sample with a luminosity range of
only 1.6\,dex, but it is more likely because we reduce the scatter in
the luminosity relationship by using only the line core component.
The open circles in Fig.~\ref{EW-L} show the integrated equivalent
widths of the \civ\ line measured directly from the original spectra
\citep{FWqc}, where \heiil\ and {O\,{\sc iii}]\,$\lambda$1664} were
also excluded.  The scatter is larger there, and the QSOs with the
largest and smallest deviations from the fit are indeed the objects
with smallest and largest weights in our SPC3 (\S\ref{Spc3},
Table~\ref{sample}).   This demonstrates that our SPC3, which
involves the broad wings of the emission lines, is the major source
of scatter in the Baldwin relationship.  The remaining scatter of
$\sim$10\% in our SPC1-derived Baldwin relationship is likely to
result from non-linearity of the  relationship(s), time variability
of spectra, and inappropriateness of L$_{\nu}$(1549) as a luminosity
indicator.  A bolometric or ionizing luminosity may be more
appropriate.  Even then, an apparent luminosity may not be that
illuminating the emission-line gas.

Besides the strongest UV lines, which are already known to show a
Baldwin effect from previous studies of direct measurements, such as
\lya, \oil, \civl, \heiil, \ciiil, and \mgiil,  SPC1 shows the
Baldwin effect in other lines.  \nvl\ appears in SPC1, showing that
its equivalent width decreases with luminosity.  This appears to
conflict with some other studies (e.g., Espey \& Andreadis 1999) that
show no Baldwin effect for \nv.  This may be because, using direct
spectral line measurements, \nv\ is difficult to deblend from \lya\
because of the uncertainties in line profiles and in determining the
location of continuum in this region.  
\nv\ and \lya\ (and some other line blends) are easily resolved in
our SPC1.  Thus SPCA provides us
with a valid way of disentangling the Baldwin effect for \nv.  We also
find a Baldwin effect for \siivoiv.   This has also been noted by
e.g., \citet{Lao95}, \citet{Gre01} although  others claim no Baldwin
effect for this feature (e.g., Cristiani \& Vio 1990; FHFC; Osmer et
al.  1994; Francis \& Koratkar 1995).

Our broad UV-optical line coverage can also link Baldwin
relationships for the UV lines with luminosity-dependent
emission-line properties in the optical region.  Our SPC1 shows that
there is a Baldwin effect for the Balmer lines, and also for broad
\heiilo.   A Baldwin effect has been seen for \hb\ by comparing the
low and high redshift data of BG92 and \citet{McI99} (see Yuan et al.
2002), and an \hb\  Baldwin effect is included in the results of
\citet{Esp99}.  However, \citet{Cro02} find a positive
correlation between EW(\hb) and luminosity. 
The reason why the Balmer-line Baldwin effect has
been difficult to identify in the past, is probably because it has
been masked, especially in integrated line measurements, by the
scatter caused by BG\,PC1 in low redshift, optical samples.  BG92
have shown an inverse luminosity dependence for the strength of broad
\heiilo\ (PC2 in their Table\,2), and their Figs.\,2 and 3 show how it
has been masked by QSO-to-QSO differences in the strength of
\feii(opt) emission.

\begin{figure}
\epsscale{0.5}
\plotone{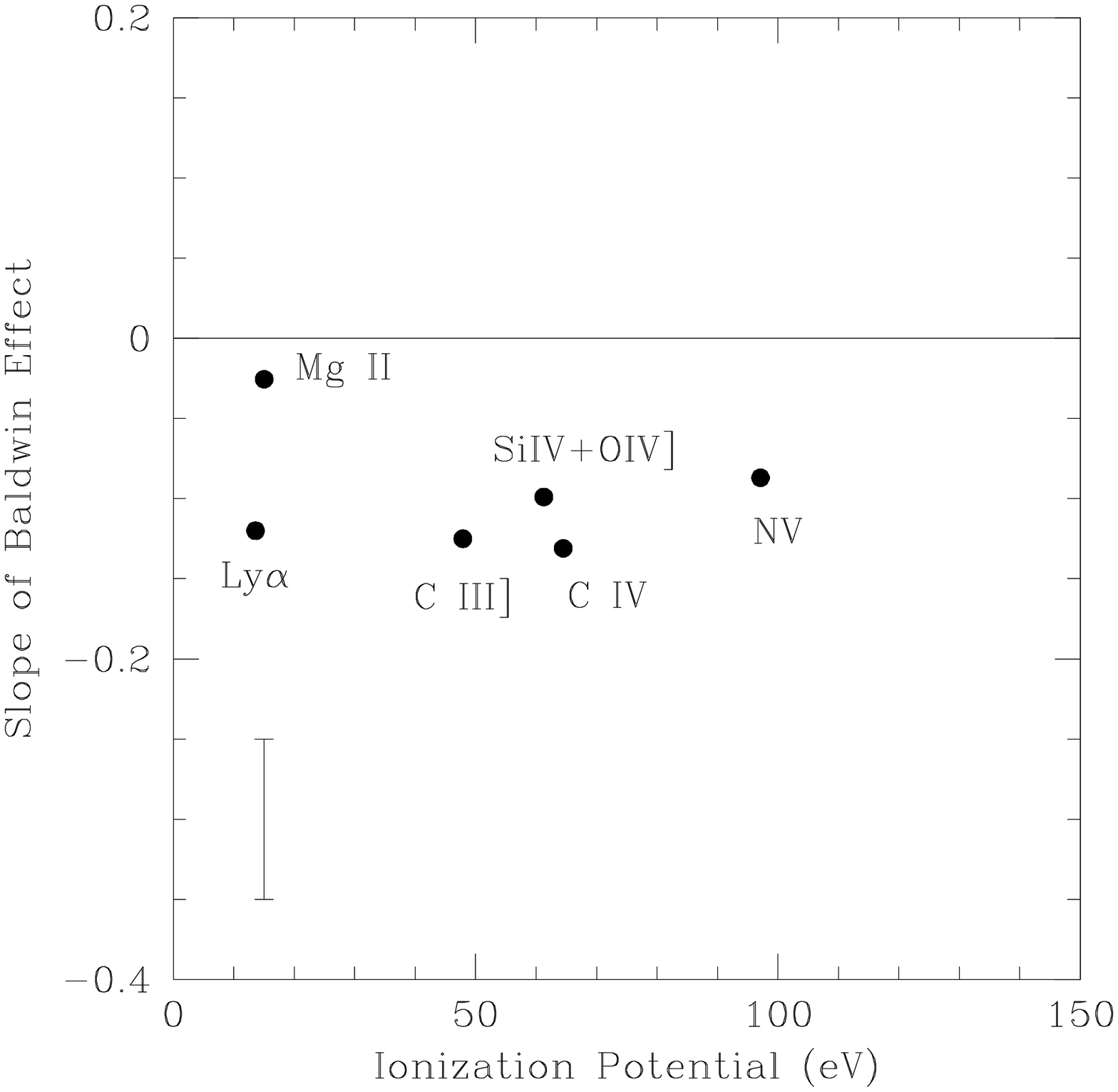}
\caption{Ionization dependence of Baldwin relationships.
The slope of the Baldwin relationship for each line is plotted 
against the ionization potential of the ion responsible (or 
mean potential in the case of \siivoiv). 
The error bar indicates the uncertainty for \civ\ from the
least-squares fitting.  The relative uncertainties from line-to-line
are much smaller and 
depend on the accuracy of the EW denominator -- the integrated line
measurement in the mean spectrum.
\label{slopes}}
\end{figure}

It has been suggested that the slope of the Baldwin effect, $\beta$,
is dependent on ionization (see Espey \& Andreadis 1999; Croom et al.
2002).  We can investigate the Baldwin effect for other broad
emission lines besides \civ.  We measure the equivalent widths of
\lya, \nv, \ciii$+$\siiii, \siivoiv, and \mgii\ the same way we do
for \civ.  We deblend \lya $+$\nv\ in the mean spectrum and measure
the narrow components of \lya\, and \nv\ in SPC1.  If indeed a linear
approximation is valid when extracting the Baldwin effect from our
SPC1, then the fractional change in integrated line equivalent width
depends  on the relative contributions from SPC1 and the mean
spectrum.  These relative contributions will determine the slopes
that we find for the Baldwin effect.  While $\beta$ for \civ\ is
uncertain to within $\pm$0.05, the slopes for different emission
lines relative to that for \civ\ can be determined quite accurately,
depending only on the uncertainty in measuring the line strength in
the mean spectrum.  The weights, W$_{\rm SPC1}$, are the same for
each emission line.  In Fig.~\ref{slopes} $\beta$\ is plotted against
the ionization potential of each ion.  The results are similar to the
dependence shown in Espey \& Andreadis' Fig.~6.  Except for \nv, they
suggest that their ``slope of slopes'' plot is consistent with
predictions by \citet{Kor98} (also shown in their Fig.~6), based on a
locally optimally emitting cloud distribution.  Our \nv\ is more
consistent with these predictions.

In summary, our broad UV-optical line coverage links the
luminosity-dependent BG\,PC2 with the Baldwin effect seen in large
samples of UV spectra, covering a wide luminosity range.  The
arguments are as follows.  Our SPC1 shows a dependence on luminosity,
consistent with Baldwin relationships seen in these larger samples.
The line cores of our SPC1 are like those seen in the FHFC UV
SPC1, for which FHFC, and \citet{Fra95} have clearly demonstrated a
Baldwin effect.  Our SPC1 extends to the optical, showing a
luminosity dependence for the Balmer lines and the broad \heiilo.
BG92's luminosity eigenvector, BG\,PC2, involves the Balmer line
equivalent widths and broad \heiilo\ (B02).  We therefore identify
BG\,PC2 with our SPC1, and
conclude that the BG\,PC2 is effectively the optical part of Baldwin
relationships derived from UV spectra.
Table~\ref{pcid} lists the identifications of the principal components
and correlated external parameters.

\begin{table}
\begin{center}
\caption{Identifications of the principal components and key
external parameters\tablenotemark{*}
\label{pcid}}
\begin{tabular}{ll}
\tableline\tableline
\multicolumn{1}{l}{Component}  & 
	\multicolumn{1}{l}{Related parameters} \\
\tableline
SPC1, line-core & SPC2 in Fig~\ref{optspca}, 
	BG PC2, FHFC SPC1, Baldwin effect, \lciv\\
SPC2, continuum slope & FHFC SPC2\\
SPC3, line-width & SPC1 in Fig~\ref{optspca}, BG PC1, $\alpha_x$\\

\tableline

\end{tabular}
\tablenotetext{*}{See Table~\ref{corr} for correlations.}

\end{center}
\end{table}


\subsection{The Second Principal Component: Continuum Slope}

Our second principal component (Fig.~\ref{spca}, SPC2) is almost a
pure continuum component.  It shows a slope at wavelengths less than
$\sim$4000\,\AA\ but flattens out in the optical region.
  This seems to be consistent with the continuum slope reported by
FHFC in the LBQS sample,  but covers a much broader wavelength
range.  The line features in this component are relatively weak, if
not absent, and probably arise from crosstalk from the other
components (see \S\ref{Sim}).  The continuum slope has a factor of 6
greater variation in this component than in the line-core component
and, therefore, accounts for most of the continuum slope variation
in the sample.  The SPC2 weights, hence continuum slopes, are
independent of luminosity (Table~\ref{corr}).

The continuum slope variation could be the result of an intrinsic
variation in the QSO continuum, e.g., the big blue bump, differing
amounts of starlight from the host galaxy, or different amounts of
reddening by dust associated with the QSO.  Any one of these possible
contributions could be orientation dependent.  We have not yet corrected
our spectra for known contributions from the host galaxy, and this is
significant at longer wavelengths for a few of the QSOs.  It is
likely that dust absorption plays an important role, as we find a
correlation between optical-UV continuum reddening,  intrinsic UV
absorption lines, and flatter soft X-ray slope \citep{Wil00,Bra00}.
If dust were the main factor producing the continuum slope variation,
we can expect it to be external to the  BLR because it would obscure
the line and continuum emission equally and hence leave the
equivalent width unchanged --- a featureless principal component, as
seen.

\subsection{The Third Principal Component: Line-width Relationships \label{Spc3}}

\begin{figure}
\epsscale{0.5}
\plotone{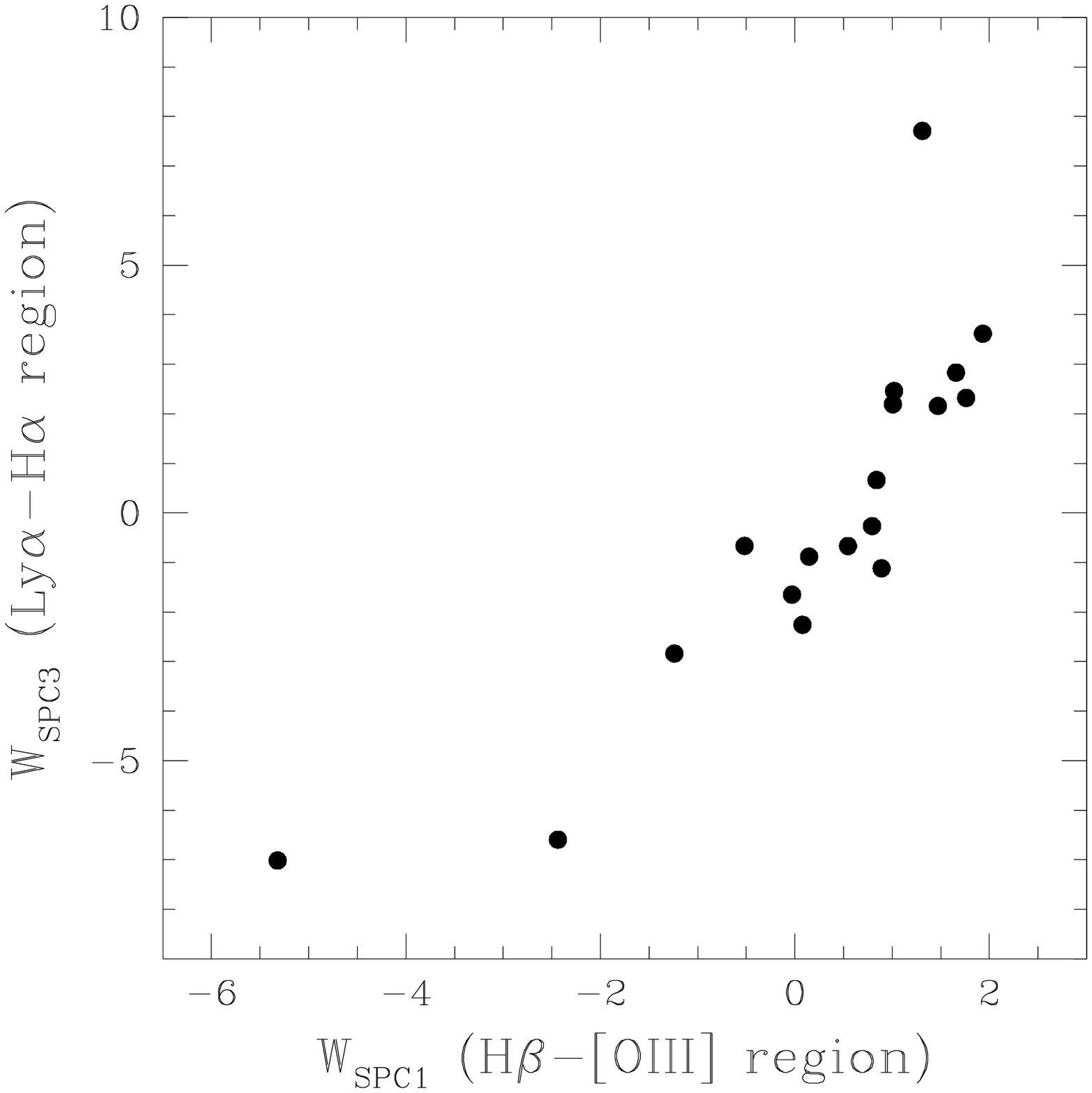}
\caption{
W$_{\rm SPC3}$ for the \lya\ -- \ha\ region vs. W$_{\rm SPC1}$ for
the \hb\ -- \oiii\
region.  The W$_{\rm SPC3}$ are from the SPCA of the UV-optical spectra
of 18 QSOs illustrated in Fig.~\ref{spca}.  The W$_{\rm SPC1}$
are for 18 QSOs of the SPCA of 22 QSO spectra of the \hb\ -- \oiii\ region 
illustrated in Fig.~\ref{optspca}.
It shows that the two principal components 
from the two separate SPCA are closely related.  
\label{ev1comp}}
\end{figure}

\begin{figure}
\epsscale{1}
\plottwo{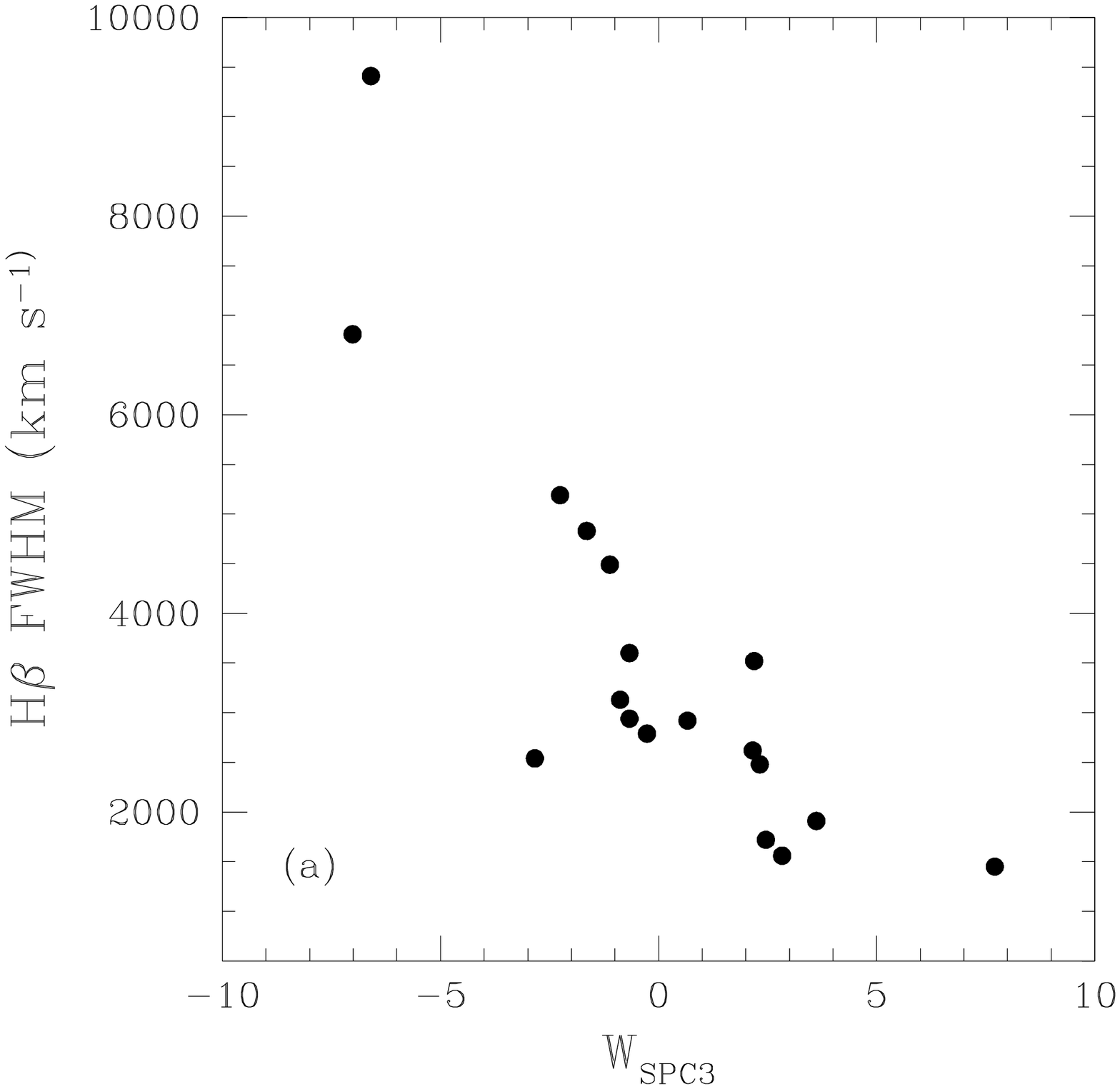}{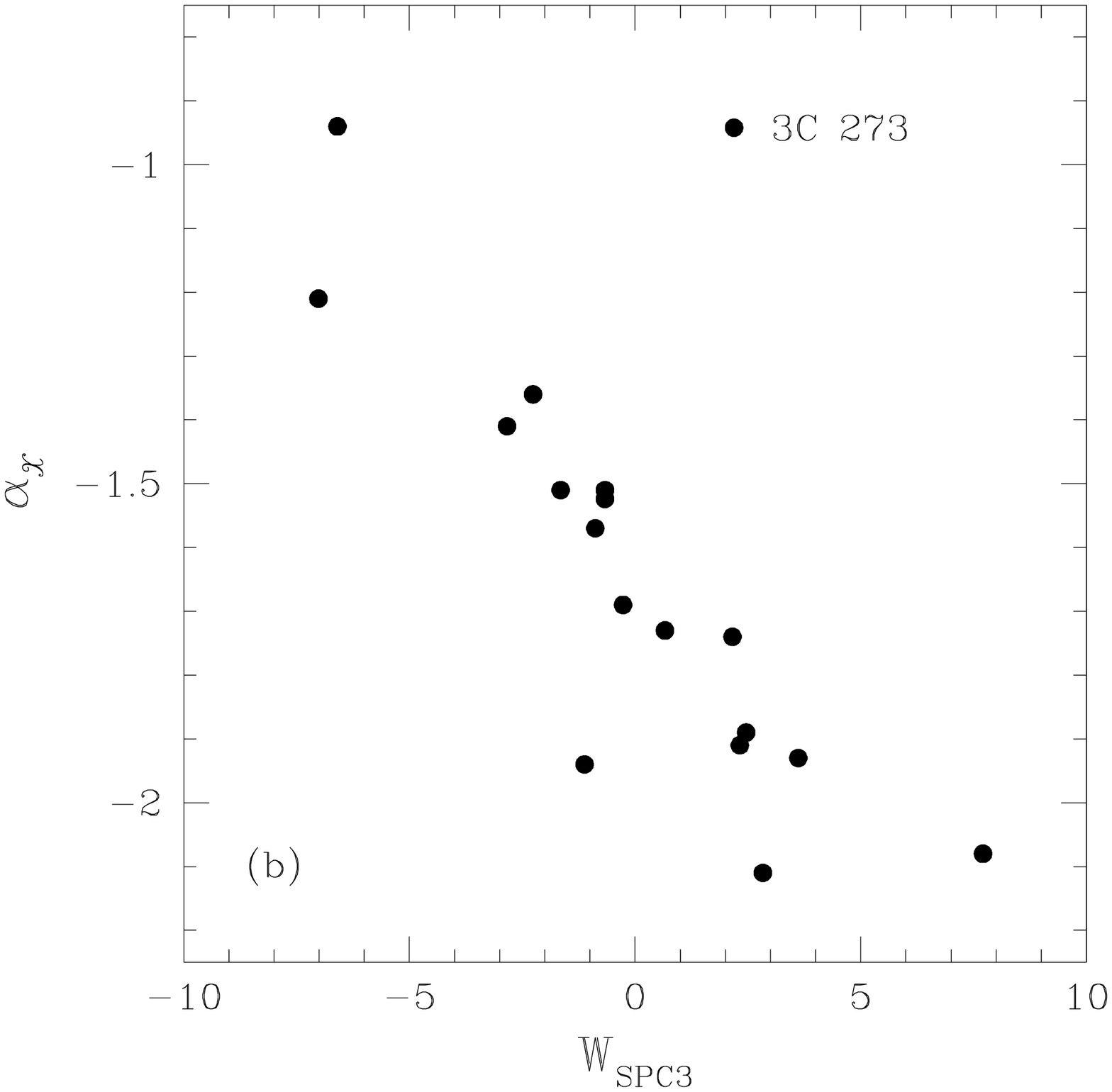}
\caption{\hb\ FWHM  and $\alpha_{x}$ vs. W$_{\rm SPC3}$ from the SPCA of
the UV-optical spectra of 18 QSOs.  Note that the soft X-ray spectrum
of 3C 273 has a significant jet contribution, and is time variable (L94).
\label{spc3-FWHM}}
\end{figure}

The \hb\ -- \oiii\ region of the third principal component (SPC3 in
Fig.~\ref{spca}) in our analysis of the entire UV-optical spectrum
is very similar to that of the SPCA of the \hb\ -- \oiii\ region
alone (SPC1 in Fig.~\ref{optspca}).  Fig.~\ref{ev1comp} compares the
weights of the two principal components and shows quantitatively that
these components are closely related.  
In SPC3, there is a clear
anticorrelation between the strengths of \oiii\ and \feii(opt)
emission and between \hb\ width and the strength of \feii(opt).
Therefore, BG\,PC1 relationships are an important part of the third
principal component in our analysis of the UV-optical spectrum.  A
prominent feature is the correlated line widths of the low ionization
broad emission lines -- \hb, \mgii\ and \ha\ (the ``W'' shape, see
\S\ref{Sim}).  Deblending of the $\lambda$1909 feature in this sample
shows that the width of \ciii\ also varies like \hb\ \citep{Wil00}.
Important variables in the BG\,PC1 eigenvector are FWHM(\hb) and the
soft X-ray spectral index $\alpha_x$ (BG92, L94, L97).  Our W$_{\rm SPC3}$ is
directly correlated with FWHM(\hb) and also with $\alpha_x$
(Fig.~\ref{spc3-FWHM}a, b, Table~\ref{corr}).

The SPC3 \civ\ feature extends over more than 30\,000\,\kms.
\civl\ does not show an obvious correlated line-width change in
SPC3.  The characteristic ``W'' shape may not appear, though, if the
line strength increases with line width.  For integrated line
measurements it is also the case that FWHM(\civ) does not correlate
well with the widths of the low-ionization lines in this sample
(Fig.~4 of Wills et al.  2000), but in these cases, the SPC1 line
core contributes some scatter to the direct measurements of FWHM
(contributing to the inverse correlation between EW(\civ) and
FWHM(\civ), Brotherton et al 1994b, \S\ref{Disc}).

As expected, the \feii (opt) features centered near 4570\,\AA\ and
5250\,\AA\ are positively correlated with \feii\ optical multiplets
27 and 28, prominent features measured at 4178\,\AA, 4237\,\AA, and 4308\,\AA\
\citep{Phi77,Phi78}.  Along with increasing strength of \feii (opt),
several low-ionization lines increase in strength:  Si\,{\sc
ii}$\lambda$1263, O\,{\sc i}$\lambda$1304, C\,{\sc ii}$\lambda$1335,
N\,{\sc iii]}$\lambda$1750, and probably Na\,{\sc i}$\lambda$5892 (for
the last, see Thompson 1991).  \siivoiv\ is also positively
correlated with \feii (opt).  The strengths of the prominent broad
emission lines correlate together -- \lya, \civ, \mgii, \hb, \& \ha,
but are anticorrelated with the low ionization lines mentioned
above.

Another important new result revealed by SPC3 is the strong
anticorrelation between broad UV \feii\ emission blends (the small
blue bump) and the optical \feii\ blends.  This was actually
predicted by \citet{Net83}.  They showed that strong \feii (opt)
requires high optical depth in the UV resonance transitions of
\feii.  At the same time, these high optical depths result in
destruction of UV \feii\ photons, in particular, by ionization from
the n=2 level of H\,{\sc i} (Balmer continuum).  
\citet{Gre98} found from composite spectra of samples
of steep and flat optical-X-ray spectral index,
$\alpha_{ox}$,
that UV and optical \feii\ emission are correlated in opposite senses
with $\alpha_{ox}$.  This suggests the anti-correlation
between UV and optical \feii\ blends found here.

\begin{figure}
\epsscale{0.5}
\plotone{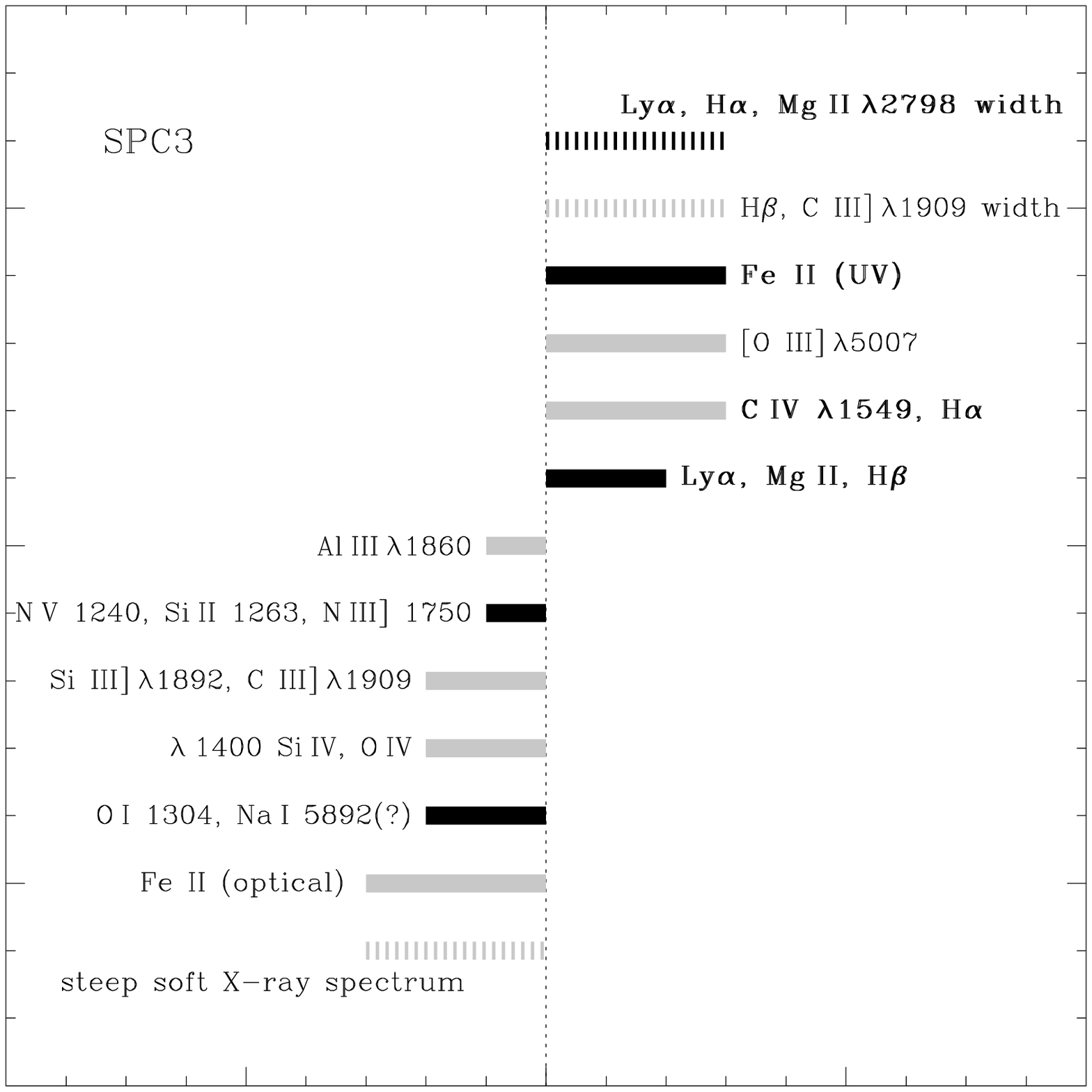}
\caption{A descriptive summary of correlations related to the 
line-width component.
Variables on the same side correlate positively with each other, and
on opposite sides, negatively.  Black bars show new correlations
found in this study.  Solid bars are for line strengths, and dotted
bars are for line width and other parameters.  
The lengths of the bars indicate
qualitatively the strength of the correlation.
\label{ev1}}
\end{figure}

SPC3 reveals most of the UV-optical BG\,PC1 relationships
demonstrated by direct emission-line and continuum measurements
(Wills et al. 1999a).  A descriptive summary of the relationships
among UV-optical parameters in SPC3 is shown in Fig.~\ref{ev1}, which
is related to Fig.\,4 in \citet{Wil99b}.  This component has a flat
continuum (SPC3 in Fig.~\ref{spca}), so changes in
UV-optical continuum shape are not obviously related to the SPC3
correlations.

\subsection{Exclusion of absorption-line QSOs and PG1202+281 \label{Bal}}

Four QSOs were excluded from the above analysis: PG\,1202+281,
because of its extremely large equivalent widths, and the three
strong absorption-line QSOs (PG\,1001+054, PG\,1114+445 and
PG\,1411+442), because their \lya\ and \civ\ absorption features
occur at different outflow velocities and thus introduce spurious
features in the principal component spectra.

To investigate the effect of 
the absorption-line QSOs, we have repeated the SPCA both
excluding the affected wavelengths, and interpolating over the absorption
features.  The results are essentially unchanged from those 
discussed above.

\begin{figure}
\epsscale{.95}
\plotone{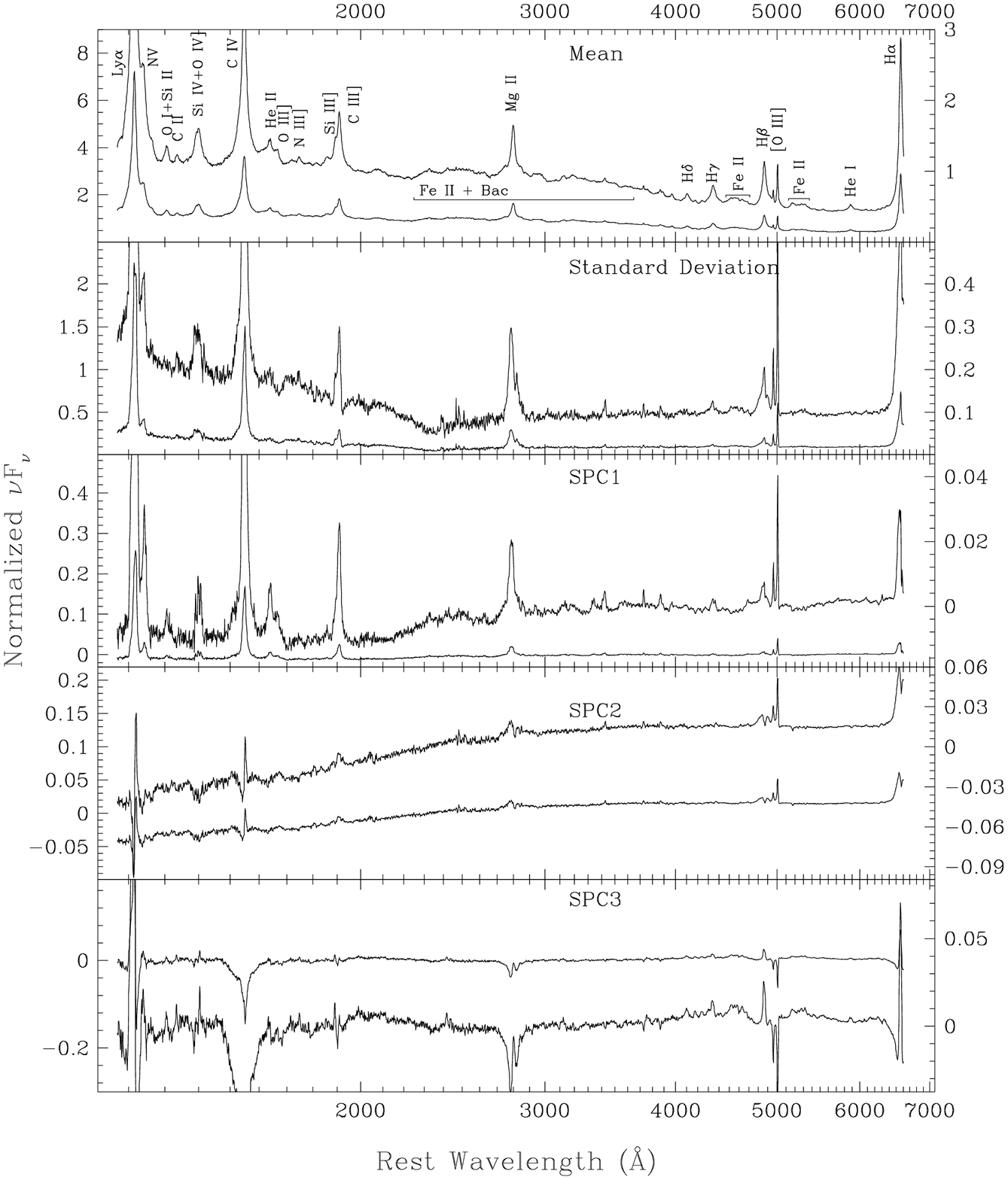}
\vspace{-15mm}
\caption{
SPCA results for the UV-optical spectra of all 22 QSOs, i.e., including
PG\,1202+281 and the absorption-line QSOs, where the absorption
regions are interpolated. 
Compare this figure with Fig.~\ref{spca} that excludes these QSOs.  
\label{spcabal}}
\end{figure}

Including PG\,1202+281, with or without the three absorption-line
QSOs, does not substantially change the results (see
Fig.~\ref{spc1-L}, Fig.~\ref{ev1comp}).  This is seen by comparing
the SPCA for the whole sample of 22 QSOs (Fig.~\ref{spcabal}) with
the adopted SPCA in Fig.~\ref{spca}.  Because of its extreme
properties, PG\,1202+281 causes a significant rotation of the
principal component axes, so the results of Fig.~\ref{spcabal} should
be viewed with caution.  The resulting crosstalk probably accounts
for the relatively strong \oiii\ emission in the SPC1 spectrum of
Fig.~\ref{spcabal}.  While PG\,1202+281 seems to be consistent with
the same principal components relationships as the other QSOs (see
Fig.~\ref{spc1-L}), its differences may hint at additional physical 
processes, or extend our parameter space to a non-linear regime.
Either way, this is a limitation of our small sample
size; a large sample covering a wider range of parameter space is
needed to address these possibilities.  In order not to give
PG\,1202+281 undue weight in the analysis, we could have carried out
SPCA using the ranks of fluxes in each wavelength bin or using
logarithmic flux scales.  For simplicity and directness of
interpretation, we have chosen not to transform the flux densities.
Therefore we have excluded PG\,1202+281.

\subsection{Simulations \label{Sim}}

\begin{figure}
\epsscale{0.6}
\plotone{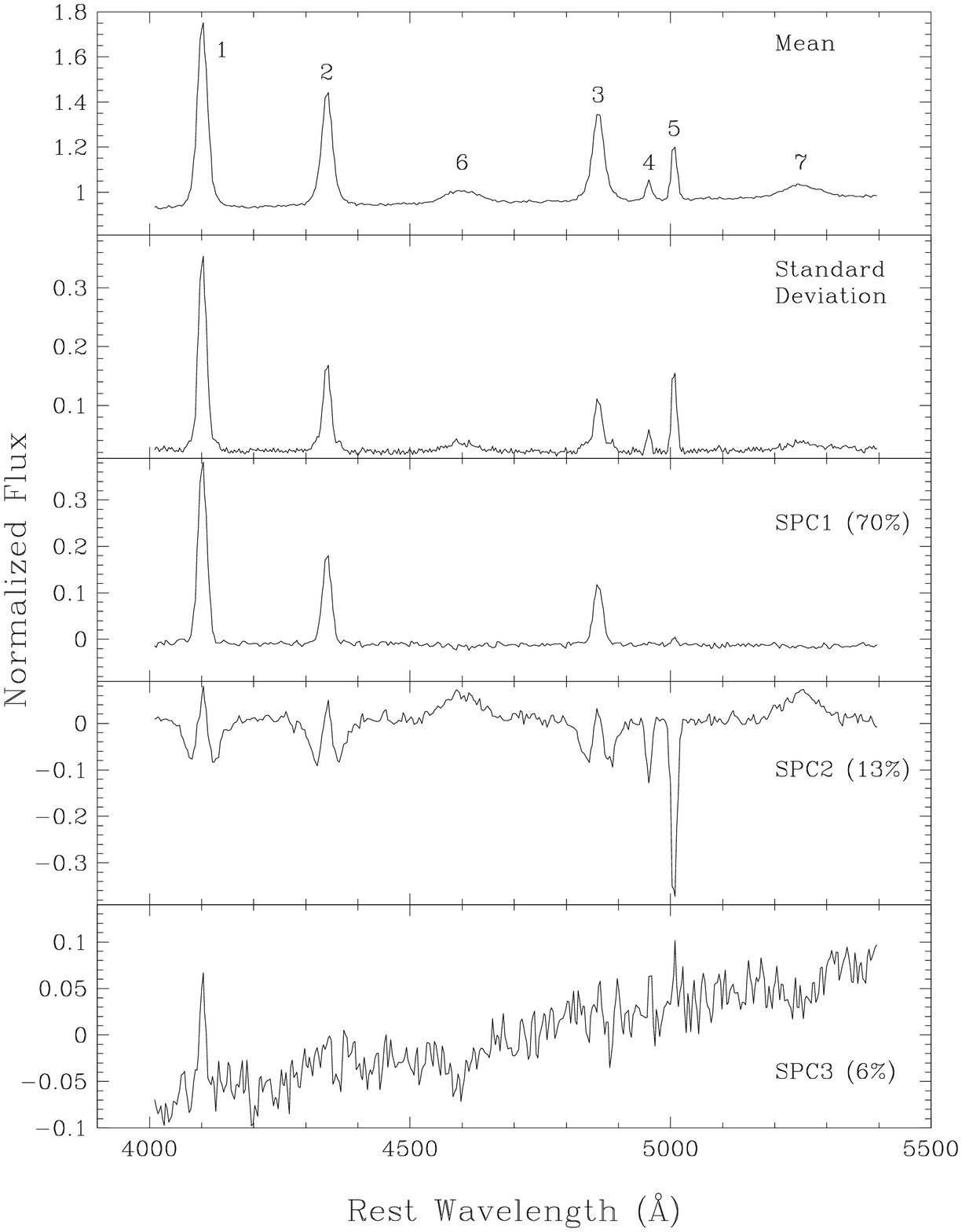}
\caption{SPCA simulation with Gaussian profile.  
Three sets of relationships input
via simulated spectra are separated into three principal 
components with
small crosstalk (e.g., lines 1 and 5 in SPC3).  Notice the ``W''
shape of lines 1, 2, and 3 in PC2.  This indicates increasing width
with increasing strength of lines 4 and 5.
The numbers in parentheses represent the percentage of the intrinsic
variance accounted for by each principal component.
\label{simgauss}}
\end{figure}

The principal components arising from PCA can be directly
interpreted only if the input parameters (i.e. the fluxes in all the bins in
this study) are related linearly.   In our SPCA, variable
line width will usually introduce non-linear relationships among the
affected binned flux densities, i.e., the flux densities do not
change linearly in different bins when the line width changes.  As an
aid to interpretation we have therefore made simulations with
artificial data (see also Mittaz, Penston, \& Snijders 1990).
Fig.~\ref{simgauss} shows an example in which we apply SPCA to 40
spectra with added Poisson noise.  Each spectrum has 7 line features
and a continuum slope, with possible relationships as indicated by
the real data.  Lines 1, 2 and 3 have one narrow and one broad
Gaussian component.  Three independent sets of relationships among
lines and continuum are included in the spectra:

\begin{enumerate}
\item{The narrow components of lines 1, 2 and 3 have a fixed width;
their strengths vary in proportion to each other.}
\item{
The broad components of lines 1, 2 and 3 have a fixed strength,
but their widths are proportional to the strength of lines 4 and 5.
The strengths of lines 4 and 5 are also proportional, and they are
inversely proportional to the strengths of lines 6 and 7.
}
\item{
Each spectrum has a different linear continuum slope,
which is not related to any line feature.   
}
\end{enumerate}

The simulation results clearly show that: 
\begin{enumerate}
\item{
The three sets of relationships are separated into the three
principal components in the order expected.
}
\item{
There is crosstalk among the three principal components, which is caused by
the non-linear effect of line width change and the continuum slope
(see \S\ref{Disc}), for example, lines 1, 2 and 5 in principal component 3 (the
slope), and line 5 in principal component 1.
(The change in continuum level changes the line equivalent widths.
But this is a real, linear effect, and not crosstalk.)
}
\item{
The line width change produces a ``W'' shape in the principal
component (or an ``M'' shape if the y axis flips), as can be seen in
principal component 2 for the broad components of lines 1, 2, and 3.
This can be visualized as the changes in the line wings.
}

\end{enumerate}

These simulation results show that SPCA is able to
successfully distinguish the relationships that we input.

\begin{figure}
\plotone{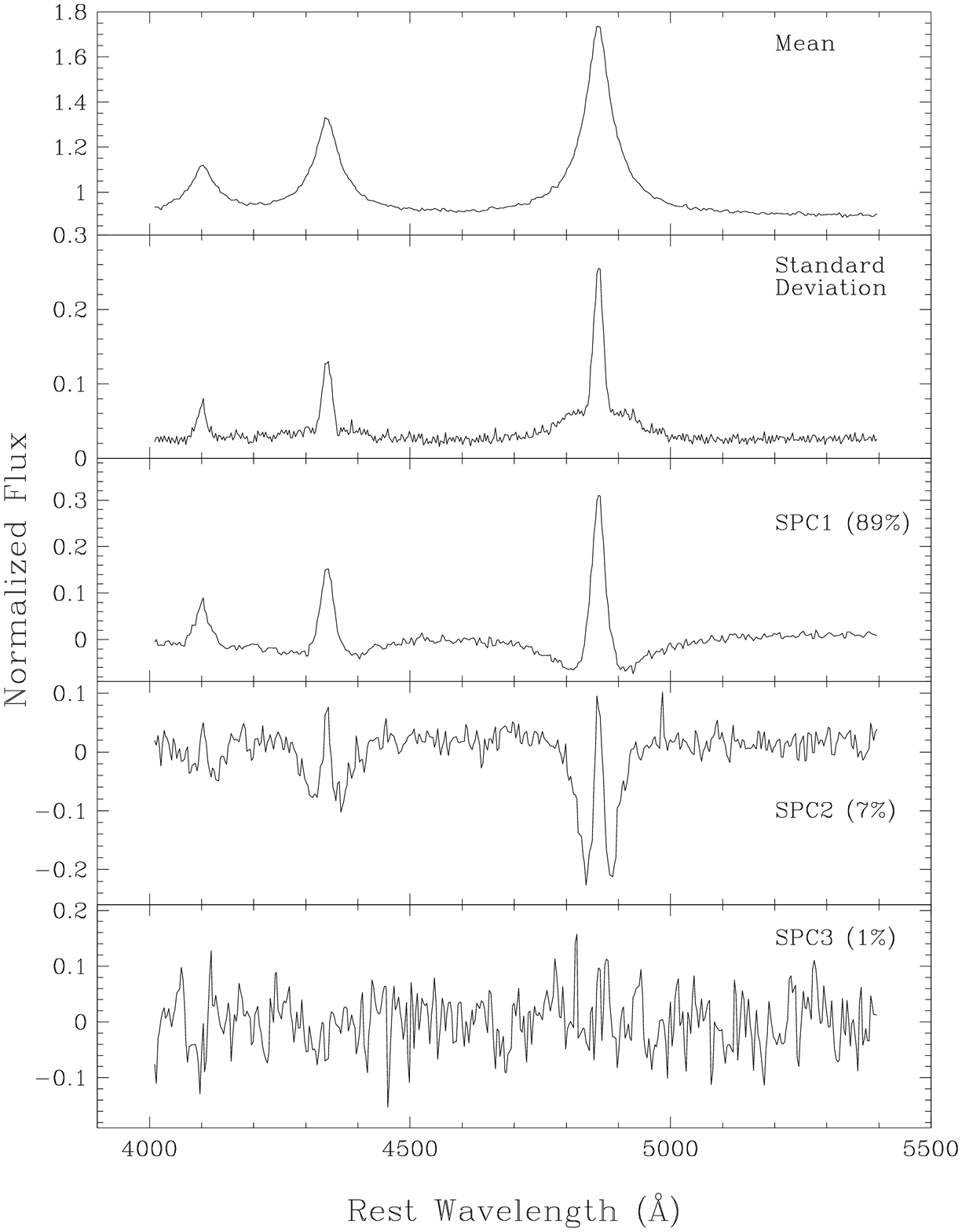}

\caption{SPCA simulation with Lorentzian profiles.  The non-linear
relationship FWHM~$\propto 1/I_p - c$ produces two dominant principal
components.  SPC1 shows that line width decreases with increasing
line strength.  SPC2 shows the effect of non-linear relationship
among the flux bins as line width changes.  The numbers in parentheses
represent the percentage of the intrinsic variance accounted for by
each principal component.
\label{simlor}}
\end{figure}

Is our interpretation really unique?  Could a non-linear relationship
produce two principal components like our SPC1 (line-core) and SPC3
(line-width)?  In order to test this, we did a simulation with three
Lorentzian line profiles for which FWHM~$\propto 1/I_p - c$, where
$I_P$ is the peak flux of the lines, $c$ is a constant chosen to vary
FWHM by a factor of 6, while $I_p$ varies by a factor of 3.  These
ranges reproduce differences seen in our real spectra.  The results
of SPCA are shown in Fig.~\ref{simlor}.  The standard deviation
spectrum shows a peak on a broad base, somewhat similar to \hb, \ha,
and \civ\ in our real data.  In the simulated SPC1, weak line wings
anticorrelate with a line core.  This is the major part of the
relationship we input, and is produced by the Lorentzian's sharp peak
and very broad wings.  This is qualitatively seen in the real SPC1
for \ha, \civ, and possibly \hb.  In the simulated SPC2 we see the
``W'' signature resulting from crosstalk caused by the non-linearity
introduced by changes in line-width.  As expected, SPC1 accounts for
much more intrinsic variance than the ``crosstalk'' component, SPC2.

To summarize, this Lorentzian simulation shows that, mathematically,
correlated line-width changes of profiles with a sharp peak and broad
wings could produce two principal components similar to SPC1 and SPC3
seen in our analysis of real data.  Nevertheless, the conclusions
derived from the real data would not be changed much:  The emission
from low velocity gas depends on luminosity, a completely independent
external parameter, but emission from non-line-core gas does not --
the Baldwin Effect remains.  SPC3 weights clearly correlate with
directly measured FWHM(\hb), and with the completely independent
external parameter, $\alpha_x$.  
A simple Lorentzian model does not account for the correlations
with external parameters; nor does it account for the different line ratio
in observational principal components.

\section{DISCUSSION \label{Disc}} 

Over the spectral range from \lya\ to \ha, SPCA shows that 78\% of
the intrinsic spectrum-to-spectrum variance is accounted for by just
three principal components (Table~\ref{pcfrac}).  In
Table~\ref{pcfrac} we also compare the fractional contributions to
the total intrinsic variance within the UV and near-UV--optical
wavelength ranges.  
We find that SPC1 contributes about half the variance
in the UV, similar to the FHFC result for 232 LBQS UV spectra.
Examination of SPC3 and higher-order component spectra, and
comparison with simulations, clearly shows the non-linear
contributions of spectrum-to-spectrum line-profile differences,
for example, differences caused by changing
width, probably kurtosis and asymmetry as well (see for
example, FHFC).  We therefore tabulate the sum of the variances from
SPC3 through SPC10 (noise increasingly dominates higher order
principal components), as probably representative of the line-width
component.  
In the UV,
the line-core component clearly dominates.  In the 2100\,\AA\ --
6607\,\AA\ range, the line-width relationships clearly dominate the
line-core component and have a contribution similar to that from the
continuum principal component.  The increasing dominance of the
line-width relationships is not surprising given the previous results
from SPCA (Fig.~\ref{optspca} and BG92).  It is easy to understand
why historically the \feii--\oiii--\hb-width correlations were
discovered for low-redshift QSOs in the optical region, and the
Baldwin effect in the UV spectra of high-redshift QSOs.  The
line-core, line-width and continuum sets of relationships together
appear to account for about 78\%--95\% of the spectrum-to-spectrum
intrinsic variance.

Much of the remaining intrinsic variance is likely to arise from
departures from linearity -- among the Baldwin and line-width
relationships for different emission lines,  in the description of
the continuum-to-continuum differences -- and from time-variability.
\citet{Pog92} showed that, for Seyfert galaxies, there is a
relationship between line and continuum luminosity, as an individual
Seyfert nucleus varies in time.  For example, for \civl, they showed
EW(\civ) $\propto$ L$^{-0.72}$, with a slope different from that for
the global (QSO-to-QSO) relationship with exponent of $-$0.17
(\S\ref{Spc1}, Peterson 1997).  They suggested that this intrinsic
``Baldwin relationship'' accounts for the scatter seen in the global
relationship. They also  demonstrated that additional scatter is
introduced by the the lag of emission-line response to continuum
variations.  We estimate the lag to contribute typically $<$5\% to
the variations in EW in our data (Fig.~\ref{EW-L}) judging from the
\hb\ lags and light curves presented by \citet{Kas00} and
\citet{Giv99}.  If the mechanism underlying the intrinsic Baldwin
Effect is similar to that underlying the global Baldwin Effect, this
source of scatter may be included in our SPC1 or SPC3 components.  If
the intrinsic and global Baldwin Effects are independent, then the
intrinsic Baldwin Effect will contribute to the residual $\sim$5\%
variance, but not to the relationships represented by the line-core
and line-width components.  That is, this source of variance would
not be important in Fig.~\ref{spc1-L}.

\subsection{The Physics Behind the Line-core and Line-width Components}

The simulation with Lorentzian line profiles (\S\ref{Sim},
Fig.~\ref{simlor}) shows that SPCA could mathematically produce a
line-core and a line-width component from a non-linear relationship.
One may argue that the results from the real QSO spectra suffer from
the same problem.  However, the correlations shown in
Table~\ref{corr} and in Figs.~\ref{spc1-L} and \ref{spc3-FWHM}b
show that the components derived from the real spectra
are physically meaningful based on their
distinct correlations with the external variables $\alpha_{x}$ and
L.  In \S\ref{Spc1} we have presented additional arguments  that the
line-core component represents the Baldwin relationship and is
dependent on luminosity (possibly related to accretion rate, \.{M}).
Our small sample, with a limited range in L, does not allow us to
distinguish whether the line-width components are related to black
hole mass, M$_{\rm BH} \propto {\rm FWHM}({\rm H}{\beta})^2 {\rm
L}^{1/2}$, or to Eddington accretion ratio, L/L$_{\rm Edd} \propto
{\rm L/M}_{\rm BH} \propto {\rm FWHM}({\rm H}{\beta})^{-2} {\rm
L}^{1/2}$ (Table~\ref{corr}).  However, the dependence of the SPC3
weights on $\alpha_{x}$, suggests, by analogy with Galactic
black-hole candidates, that this component is driven by Eddington
accretion ratio (L94, L97, Yuan et al. 2002).  Stronger EW \feii(opt),
\feii(opt)/\feii(UV), and the \feii (opt) -- \oiii\ anticorrelation
indicate that more dense gas is available to fuel the central
engine.  Stronger \feii(opt) is associated with narrower FWHM(\hb),
consistent with an Eddington ratio interpretation of SPC3.
Further support for a physical difference between SPC1 and SPC3 gas is
indicated by line-intensity ratios: \citet{Bro94a} showed that the
emission-line ratios for gas of low- and high-velocity dispersion
indicate different photoionizing flux, ionization state, and
density.

The near-absence of emission in SPC1 from gas with high-velocity
dispersion, and the luminosity independence of SPC3, suggest
that the line emission represented by SPC3 has EW $\sim$ constant.
Could the global Baldwin relationship be explained  by a very simple
model in which a total line equivalent width is the sum of the
equivalent width of the line-width component ($\sim$ constant), and
that of the line-core component ($\propto$ L$^{-1}_{\rm cont}$),
i.e., EW(total) = $a + b {\rm L}^{-1}_{\rm cont}$?  Such a dependence
can fit our own data, but cannot fit the Baldwin effect data for
large samples, such as that of \cite{Kin90}, covering  seven orders
of magnitude in luminosity.

\subsection{Separating the Line-core and Line-width Spectra}

Since the relationships represented by the line-core and line-width
spectral principal components have real, independent, physical
origins, it would
be appropriate to model these relationships independently
rather than, as in the past, trying to model the observed integrated
equivalent widths (e.g., Korista et al. 1998; Kuraszkiewicz et al.
2000).  A case in point may be attempts to ascribe an ionization-potential
dependence to $\beta$, the exponent of the Baldwin relationship.  The
{\it relative} values of $\beta$ are quite accurately determined for
different emission lines, but the scatter is large (Fig.~\ref{slopes}).
The large scatter may be attributed to the use of integrated EWs with
a significant contribution from non-core emission because the
non-core line ratios are independent of those in SPC1 gas.

One possibility is to perform a SPCA on a sample of simulated model
spectra, making a comparison of the resulting SPC spectra with those
from a SPCA of the observed spectra.  Another approach might be to
use the principal component spectra to derive separate spectra for
the (assumed separate) gas components responsible for the line-core
and line-width components, and compare each of these with
photoionization models.  
This was attempted by \citet{Bro94a} who
decomposed UV emission lines into components 
representing intermediate and 
very broad line regions (ILR and VBLR).
Our SPCA shows that  their VBLR is an over-simplification.
While their ILR is similar to our SPC1 component, we do not define a
VBLR, but rather a line-width component (dominated by our SPC3),
where line widths range from the narrowest to broadest \hb\ lines
observed (FWHM from 1500 to 9500\,\kms).  Thus the ILR derived
by  \citet{Bro94a} would include SPC1 emission as well as
low-velocity emission from the line-width component.  However, the
technique of best-fitting a combination of line-core and line-width
spectra still seems promising.  Further discussion is beyond the
scope of the present paper.

While the ILR has been directly related to the Baldwin effect
(Brotherton et al. 1994a, FHFC, Francis \& Koratkar 1995), the
relationship of the UV ILR to the optical BG\,PC1 has been
controversial, including the question of whether BG\,PC1 is part of
the Baldwin relationship, or whether it is the cause of its enormous
scatter \citep{BFqc}.  Because of our wide wavelength coverage we
have been able to demonstrate clearly that these two sets of
relationships  are orthogonal.  \citet{BFqc} came to a different
conclusion, based on their finding of increasing ILR strength with
[O\,III]$\lambda$5007, a strong SPC3 or BG\,PC1 parameter.  This may
be because of the ILR decomposition problem mentioned above.

\section{SUMMARY AND CONCLUSIONS \label{Summ}}

Our Spectral Principal Component Analysis of an essentially complete
sample of 22 PG QSOs yielded three significant linear spectral
principal components accounting for $\sim$78\% of the intrinsic variance
among emission-line and continuum spectra over the wavelength range
from \lya\ to \ha\ (excluding the three QSOs with strong intrinsic
absorption and the extreme object PG1202\,+281).  Spectral Principal
Component 1 dominates the UV emission lines, accounts for $\sim$41\% of the
intrinsic spectrum-to-spectrum variance, and is identified with the
Baldwin relationships.  Spectral Principal Component 2 accounts for
most of the continuum variations, contributing $\sim$23\% of the total
intrinsic variance.  Spectral Principal Component 3 is directly
related to the first Principal Component in the PCA of integrated
line measurements in the \feii--\hb--\oiii\ region by BG92.  It
includes line-width, which introduces non-linear relationships among
flux bins across emission lines, and therefore propagates to
higher-order Principal Components in our linear analysis.  This
line-width component dominates the \feii--\hb--\oiii\ region, and
accounts for between 14\% and 31\% of the sample's total intrinsic
variance.  Tables~\ref{pcfrac} and \ref{pcid} summarize these 
results.

\subsection{The Baldwin Effect and SPC1}
\begin{enumerate}

\item SPC1 is dominated by emission from gas of low velocity
dispersion (FWHM 2000 -- 3000\,\kms), but also includes
broad \heiilo\ (FWHM 4880\,\kms).

\item The weights of this principal component correlate with the
QSOs' luminosities.

\item Deriving the equivalent widths from this principal component
leads to a Baldwin relationship in good agreement with the slope and
normalization determined for much larger samples and luminosity
ranges spanning 7 orders of magnitude.

\item  The narrow (FWHM 2000 -- 3000\,\kms) features in our Spectral
Principal Component 1 agree in profile and line intensity ratios with
those derived from SPCA for much larger LBQS samples using just UV
spectra from \lya\ to the \ciii\ blend (Principal Component 1, from
Francis et al. 1992; Francis \& Koratkar 1995), and inferred by
comparison of mean UV spectra of low and high luminosity QSOs
\citep{Osm94}, 
and of low and high X-ray brightness QSOs divided by $\alpha_{ox}$
\citep{Gre98}.
\citet{Fra95} showed quantitative agreement with the
Baldwin effect, for their large, but low signal-to-noise ratio sample.

\item The strong, broad
\heiilo\ links our SPC1 directly with the ``luminosity'' principal
component derived by BG92 and B02 from direct measurements.

\item The Baldwin relation determined from direct equivalent width
measurements would not be expected to be very significant for a
sample the size of ours and covering only 2 orders of magnitude in
luminosity.  However we demonstrate a significant reduction in
scatter using SPCA-derived EWs, over those measured directly.  The
line-width components contribute most of the scatter in the Baldwin
relationships.  This demonstrates the potential of SPCA-derived EWs
to define a much cleaner Baldwin relation for future investigations
of emission line response to the ionizing continuum, perhaps with
applications to cosmology (relations between luminosity distance and
redshift).

\item We have determined the Baldwin relationships for different emission
lines, confirming that the effect is present for the Balmer lines.
UV and optical \feii\ emission-line blends are very weak or absent
from our ``Baldwin Effect'' principal component.

\item Our clean separation of \lya\ and \nv\ in SPC1 shows a
significant Baldwin effect for \nv.  This result is in striking
contrast with the slope derived from direct measurements, where the
EW shows a positive slope (increases) with luminosity.

\end{enumerate}

\subsection{The Line-width Components (SPC3 and higher order)}

\begin{enumerate}

\item  We have demonstrated clearly that the third principal
component is directly related to BG92's first principal component.
The SPC3 weights are shown to directly correlate with FWHM(\hb) and
with $\alpha_{x}$, parameters that are important in BG\,PC1
correlations in the \hb\ region (BG92 and L94, L97).

\item  We present a number of emission line correlations, both
positive and negative.  Of particular note is a clear anticorrelation
between optical and UV \feii\ blends, as predicted by \citet{Net83}.

\item  \citet{Laor98}, and \citet{Geb00} have demonstrated a
calibration of black hole mass in terms of FWHM \hb\ and luminosity,
as if \hb\ line width is determined by virial motions.  The
line-width correlations demonstrated by this component, between
Balmer lines and \mgii, and FWHM \ciii\ has also been shown to
correlate with \hb\ \citep{Wil99c}, suggesting that this calibration
can be used to investigate black-hole mass relationships at higher
redshifts ($z$ \simgt 2 -- 3.5), where \mgii\ and \ciii\ are
accessible to ground-based spectroscopy (see also McLure \& Jarvis
2002).

\end{enumerate}

\subsection{The Relation Between Line-width and Line-core Components}

We have demonstrated the power of our wide wavelength coverage
combined with SPCA to link two independently-discovered luminosity
relationships at high and low redshift, showing that the BG92 and B02
luminosity principal component of low redshift samples is directly
related to the Baldwin effect in higher redshift, UV spectra, and
showing that BG\,PC1 (or the line-width principal component)
relationships are the cause of enormous scatter in Baldwin
relationships derived from integrated, direct, EW measurements.  We
show how Baldwin relationships can be derived using SPCA, virtually
eliminating the scatter.  Such spectrum-luminosity relationships
might be used to determine $z$-independent luminosities, hence to
derive L-$z$ relationship at high $z$.  This rekindles the hope that
Baldwin relationships can be used for cosmology.  The line-width
principal components are related to black-hole mass or Eddington
accretion ratio, and their clean separation from the Baldwin
relationships can, in principle, lead to their use to trace QSO
evolution.  The wide wavelength coverage and SPCA also provides a
strong visual impressions of spectral relationships that provide a
guide for future direct line measurements and analysis.

\acknowledgements

We thank Paul Francis and Mike Brotherton for many discussions, and
for comments on a draft of this paper.  We appreciate Bill Jefferys'
help with the Bayesian analysis, and thank Scott Croom for a copy of
his paper in advance of publication.  We also thank the referee,
Paul Green, for his valuable suggestions and careful reading of
the paper.  We gratefully acknowledge the
help of C. D. (Tony) Keyes and A. Roman of STScI, M. Dahlem (now of
ESTEC), David R. Doss, Jerry Martin, Earl Green, and Larry Crook at
McDonald Observatory, and also M. Cornell and R. Wilhelm, who
provided computer support in the Astronomy Department, and at
McDonald.  BJW acknowledges financial support by NASA through LTSA
grant NAG5-3431 and grant GO-06781 from the Space Telescope Science
Institute which is operated by the Association of Universities for
Research in Astronomy, Inc., under NASA contract NAS5-26555.  We have
made use of the NASA/IPAC Extralagactic Database (NED), which is
operated by the Jet Propulsion Laboratory, California Institute of
Technology, under contract with NASA.



\begin{thebibliography}{}

\bibitem[Alexander \& Netzer(1994)]{AleNet}
Alexander, T., \& Netzer, H. 1994, \mnras, 270, 781

\bibitem[Baldwin(1977)]{Bal77}Baldwin, J.~A.\ 1977, ApJ, 214, 679

\bibitem[Baldwin, Wampler, \& Gaskell(1989)]{Bal89} Baldwin,
 J.~A., Wampler, E.~J., \& Gaskell, C.~M.\ 1989, \apj, 338, 630

\bibitem[Baldwin et al.(1995)]{Bal95} Baldwin, J.~A., Ferland, G.,
Korista, K., \& Verner, D. 1995 \apj, 455, L119

\bibitem[Bohlin(1996)]{Boh96}Bohlin, R.~C.\ 1996, AJ, 111, 1743

\bibitem[Bohlin(2000)]{Boh00}Bohlin, R.~C.\ 2000, AJ, 120, 437

\bibitem[Boroson \& Green(1992)]{BG92}Boroson, T., \& Green,
R.~F.\ 1992, ApJS, 80, 109 (BG92)

\bibitem[Boroson(2002)]{Bor02} Boroson, T.~A.\ 2002, \apj, 565, 78
(B02)

\bibitem[Bottorff \& Ferland(2001)]{Bot01} Bottorff, M.~C., \&
Ferland, G.~J. 2001, \apj, 549, 118

\bibitem[Brandt, Laor, \& Wills(2000)]{Bra00} Brandt, W.~N., Laor,
A., \& Wills, B.~J.  2000, \apj, 528, 637

\bibitem[Brotherton et al.(1994a)]{Bro94a}Brotherton, M.~S., Wills,
B.~J., Francis, P.~J., \&
 Steidel, C.~S.\ 1994a, \apj, 430, 495

\bibitem[Brotherton et al.(1994b)]{Bro94b}
 Brotherton, M.~S., Wills, B.~J., Steidel, C.~C., \& Sargent,
 W.~L.~W.\ 1994b, \apj, 423, 131


\bibitem[Brotherton \& Francis(1999)]{BFqc}Brotherton, M.~S., \&
Francis, P.~J.\ 1999, in ASP Conf.  Series 162, Quasars and
Cosmology, ed. G.~J. Ferland, \& J.~A.\ Baldwin (San Francisco: ASP), 395


\bibitem[Cristiani \& Vio(1990)]{1990A&A...227..385C} Cristiani,
S., \& Vio, R.\ 1990, \aap, 227, 385

\bibitem[Croom et al.(2002)]{Cro02}Croom, S.~M. et al. 2002, \mnras,
337, 275


\bibitem[Espey \& Andreadis(1999)]{Esp99} Espey, B.~R., \& Andreadis,
S.\ 1999, in ASP Conf. Series 162, Quasars and Cosmology, ed. G.~J.
Ferland, \& J.~A.\ Baldwin (San Francisco: ASP), 351

\bibitem[Fabian(1999)]{Fab99} Fabian, A.~C.\ 1999, \mnras, 308, L39


\bibitem[Ferrarese \& Merritt(2000)]{FerMer00} Ferrarese, L., \&
Merritt, D.\ 2000, \apjl, 539, L9

\bibitem[Ferland \& Baldwin(1999)]{FBqc}Ferland, G.~J., \& Baldwin,
J.~A.\ 1999, ASP Conf. Series 162, Quasars and Cosmology
(San Francisco: ASP)


\bibitem[Francis et al.(1992)]{F92}Francis, P.~J., Hewett, P.~C.,
Foltz, C.~B., \& Chaffee, F.~H.\ 1992, ApJ, 398, 476  (FHFC)

\bibitem[Francis \& Wills(1999)]{FWqc}Francis, P.~J., \& Wills,
B.~J.\ 1999, in ASP Conf. Series 162, Quasars and Cosmology, ed.
G.~J. Ferland, \& J.~A. Baldwin (San Francisco: ASP), 373

\bibitem[Francis \& Koratkar(1995)]{Fra95} Francis, P.~J., \&
Koratkar, A.\ 1995, \mnras, 274, 504

\bibitem[Gebhardt et al.(2000)]{Geb00} Gebhardt, K.~et al.\ 2000,
\apjl, 543, L5


\bibitem[Gilli, Salvati, \& Hasinger(2001)]{Gil01} Gilli, R.,
Salvati, M., \& Hasinger, G.\ 2001, \aap, 366, 407

\bibitem[Giveon et al.(1999)]{Giv99} Giveon, U., Maoz, D., Kaspi, S.,
Netzer, H., \& Smith, P.~S.\ 1999, \mnras, 306, 637

\bibitem[Goldschmidt et al.(1992)]{Gol92} Goldschmidt, P., Miller, L.,
La Franca, F., \& Cristiani, S. 1992, \mnras, 256, 65

\bibitem[Green(1998)]{Gre98} Green, P.~J.\ 1998, \apj, 498, 170

\bibitem[Green, Forster, \& Kuraszkiewicz(2001)]{Gre01} Green, P.~J.,
Forster, K., \& Kuraszkiewicz, J.\ 2001, \apj, 556, 727


\bibitem[Kaspi et al.(2000)]{Kas00} Kaspi, S., Smith, P.~S., Netzer,
H., Maoz, D., Jannuzi, B.~T., \& Giveon, U.\ 2000, \apj, 533, 631


\bibitem[Kinney et al.(1990)]{Kin90}Kinney, A.~L., Rivolo, A.~R., \&
Koratkar, A.~P.\ 1990, \apj, 357, 338

\bibitem[Korista, Baldwin, \& Ferland(1998)]{Kor98}Korista, K.~T.,
Baldwin, J.~A., \& Ferland, G.~J.\ 1998, \apj, 507, 24

\bibitem[Korista(1999)]{Kor99}Korista, K.~T. 1999, in ASP Conf.
Series 162, Quasars and Cosmology, ed.  G.~J. Ferland, \& J.~A.
Baldwin (San Francisco: ASP), 429

\bibitem[Kormendy \& Ho(2000)]{Kor00} Kormendy, J., \& Ho,
L.~C.\ 2000, to appear in The Encyclopedia of Astronomy and
Astrophysics (Institute of Physics Publishing), astro-ph/0003268

\bibitem[Kuraszkiewicz, Wilkes, Czerny, \& Mathur(2000)]{Kur00}
Kuraszkiewicz, J., Wilkes, B.~J., Czerny, B., \& Mathur, S.\ 2000,
\apj, 542, 692

\bibitem[Laor et al.(1994)]{L94a}Laor, A., Fiore, F., Elvis, M.,
Wilkes, B.~J., \& McDowell, J.~C.\ 1994, \apj, 435, 611 (L94)


\bibitem[Laor et al.(1995)]{Lao95}Laor, A., Bahcall, J.~N., Jannuzi,
B.~T., Schneider, D.~P., \& Green, R.~F.\ 1995, \apjs, 99, 1

\bibitem[Laor et al.(1997)]{Lao97} Laor, A., Fiore, F., Elvis, M.,
Wilkes, B.~J., \& McDowell, J.~C.\ 1997, \apj, 477, 93 (L97)

\bibitem[Laor(1998)]{Laor98}Laor, A.\ 1998, ApJ, 505, L83

\bibitem[Marconi \& Salvati(2002)]{Mar02}  Marconi, A., \& Salvati,
M.\ 2002, in ASP Conf. Proc. 258, Issues in Unification of AGNs, ed. 
R.~Maiolino, A.~Marconi, \& N.~Nagar (San Francisco: ASP), 217

\bibitem[McIntosh et al.(1999)]{McI99} McIntosh, D.~H., Rieke, M.~J.,
Rix, H.-W., Foltz, C.~B., \& Weymann, R.~J.\ 1999, \apj, 514, 40

\bibitem[McLure \& Jarvis(2002)]{McL02} McLure, R.~J., \& Jarvis,
M.~J. 2002, \mnras, 337, 109

\bibitem[Mickaelian et al.(2001)]{Mic01} Mickaelian, A.~M., Gon\c{c}alves,
A.~C., V\'{e}ron-Cetty, M.~P., \& V\'{e}ron, P. 2001, Astrophysics, 44, 14

\bibitem[Mittaz, Penston, \& Snijders(1990)]{Mit90} Mittaz, J.~P.~D.,
Penston, M.~V., \& Snijders, M.~A.~J.\ 1990, \mnras, 242, 370

\bibitem[Murray \& Chiang(1997)]{MC97} Murray, N., \& Chiang,
J.\ 1997, \apj, 474, 91



\bibitem[Netzer \& Wills(1983)]{Net83} Netzer, H., \& Wills,
B.~J.\ 1983, \apj, 275, 445

\bibitem[Netzer \& Peterson(1997)]{Net97} Netzer, H., \& Peterson,
B.~M.\ 1997, in Astronomical Time Series, ed. D. Maoz, A. Sternbery,
\& E. Leibowitz (Dordrecht: Kluwer), 85

\bibitem[Osmer et al.(1994)]{Osm94}Osmer, P.~S., Porter, A.~C., \&
Green, R.~R.\ 1994, \apj, 436, 678

\bibitem[Peterson(1997)]{Pet97}Peterson, B.~M.\ 1997, An Introduction
to Active Galactic Nuclei, p91 (United Kingdom:Cambridge University Press)

\bibitem[Phillips(1977)]{Phi77} Phillips, M.~M.\ 1977, \apj, 215, 746

\bibitem[Phillips(1978)]{Phi78} Phillips, M.~M.\ 1978, \apj, 226, 736

\bibitem[Pogge \& Peterson(1992)]{Pog92} Pogge, R.~W., \& Peterson,
B.~M.\ 1992, \aj, 103, 1084

\bibitem[Pounds, Done, \& Osborne(1995)]{PDO95} Pounds, K.~A., Done,
C., \& Osborne, J.~P. 1995, \mnras, 277, L5

\bibitem[Predehl \& Schmitt(1995)]{Pre95} Predehl, P., \& Schmitt,
J.~H.~M.~M.\ 1995, \aap, 293, 889


\bibitem[Schmidt \& Green(1983)]{Sch83}Schmidt, M., \& Green,
R.~F.\ 1983, \apj, 269, 352

\bibitem[Scoville \& Norman(1998)]{SN98} Scoville, N., \& Norman, C.
\apj, 332, 163

\bibitem[Silk \& Rees(1998)]{Sil98}Silk, J., \& Rees, M.~J.\ 1998,
\aap, 331, L1

\bibitem[Thompson(1991)]{Tho91}Thompson, K.~L.\ 1991, \apj, 374, 496


\bibitem[Wampler \& Ponz(1985)]{WamPon85} Wampler, E.~J. \& Ponz, D.
1985, \apj, 298, 448

\bibitem[Wandel, Peterson, \& Malkan(1999)]{Wan99} Wandel, A.,
Peterson, B.~M., \& Malkan, M.~A.\ 1999, \apj, 526, 579




\bibitem[Wills et al.(1999a)]{Wil99a} Wills, B.~J., Laor, A.,
Brotherton, M.~S., Wills, D., Ferland, G.~J., \& Shang,
Zhaohui\ 1999a, ApJ, 515, L53

\bibitem[Wills et al.(1999b)]{Wil99b} Wills, B.~J., Brotherton,
M.~S., Laor, A., Wills, D., Wilkes, B.~J., Ferland, G.~J., \& Shang,
Zhaohui\  1999b, in ASP Conf. Series 162, Quasars and Cosmology, ed.
G.~J. Ferland, \& J.~A. Baldwin (San Francisco: ASP), 373

\bibitem[Wills et al.(1999)]{Wil99c} Wills, B.~J., Brotherton, M.~S.,
Laor, A., Wills, D., Wilkes, B.~J., \& Ferland, G.~J.\ 1999c, in ASP
Conf.~Ser.~175, Structure and Kinematics of Quasar Broad Line
Regions, ed. C. M. Gaskell, W. N. Brandt, M.  Dietrich, D.
Dultzin-Hacyan, \& M. Eracleous (San Francisco: ASP), 241

\bibitem[Wills, Shang, \& Yuan(2000)]{Wil00} Wills, B.~J., Shang,
Zhaohui, \& Yuan, Juntao\ 2000, New Astronomy Reviews, 44, 511

\bibitem[Yuan \& Wills(2002)]{Yua02} Yuan, Juntao, \& Wills,
B.~J.\ 2002, in preparation

\end{thebibliography}
\end{document}